\documentclass[12pt]{article}

\usepackage[centertags]{amsmath}
\usepackage{amsmath}
\usepackage{amsfonts}
\usepackage{amssymb}
\usepackage{amsthm}
\usepackage{newlfont}
\usepackage{amscd}
\usepackage{empheq}
\usepackage{epsfig,graphics,graphicx,amsfonts,psfrag}
\usepackage{verbatim}
\usepackage{eucal}

\newcommand{\NN}{{\mathbb N}}

\newcommand{\RR}{{\mathbb R}}

\newcommand{\beq}{\begin{equation}}
\newcommand{\eeq}{\end{equation}}
\newcommand{\ba}{\begin{array}}
\newcommand{\ea}{\end{array}}
\newcommand{\bea}{\begin{eqnarray}}
\newcommand{\eea}{\end{eqnarray}}
\newcommand{\eps}{{\epsilon}}

\usepackage{graphicx}
\usepackage[usenames,dvipsnames]{color}
\usepackage{transparent} 

\numberwithin{equation}{section}
\allowdisplaybreaks

\begin{document}

\begin{center}
{\large \sc \bf On the dispersionless Kadomtsev-Petviashvili equation with arbitrary nonlinearity and dimensionality:\\ exact solutions, longtime asymptotics of the Cauchy problem, wave breaking and discontinuous shocks}

\vskip 10pt

{\large  F. Santucci$^{1,3}$ and P. M. Santini$^{2,4}$}

\vskip 10pt

{\it 
$^1$ Istituto Superiore IIS G.Marconi,  via Reno snc, 04100 Latina, Italy

\smallskip

$^2$ Dipartimento di Fisica, Universit\`a di Roma "La Sapienza", and \\
Istituto Nazionale di Fisica Nucleare, Sezione di Roma 1 \\
Piazz.le Aldo Moro 2, I-00185 Roma, Italy}

\vskip 10pt

$^{3}$e-mail:  {\tt nadir17@libero.it}\\
$^{4}$e-mail:  {\tt paolo.santini@roma1.infn.it}
\vskip 10pt

{\today}

\end{center}

\begin{abstract}
We study the generalization of the dispersionless Kadomtsev - Petviashvili (dKP) equation in $n+1$ dimensions and with 
nonlinearity of degree $m+1$, a model equation describing the propagation of weakly nonlinear, quasi one dimensional waves in the absence of dispersion and dissipation, and arising in several physical contexts, like acoustics, plasma physics, hydrodynamics and nonlinear optics. In $2+1$ dimensions and with quadratic nonlinearity, this equation is integrable through a novel IST, and it has been recently shown to be a prototype model equation in the description of the 
two dimensional wave breaking of localized initial data. In higher dimensions and with higher nonlinearity, the generalized dKP equations are not integrable, but their invariance under motions on the paraboloid allows one to construct in this paper a family of exact solutions  describing waves constant on their paraboloidal wave front and breaking simultaneously in all points of it, developing after breaking either multivaluedness or  single valued discontinuous shocks. Then such exact solutions are used to build the longtime behavior of the solutions of the Cauchy problem, for small and localized initial data, showing that wave breaking of small initial data takes place in the longtime regime if and only if $m(n-1)\le 2$. At last, the analytic aspects of such a wave breaking are investigated in detail in terms of the small initial data, in both cases in which the solution becomes multivalued after breaking or it develops a discontinuous shock. These results, contained in the 2012 master thesis of one of the authors (FS) \cite{thesis}, generalize those obtained in \cite{MS_dkpn} for the dKP equation in $n+1$ dimensions with quadratic nonlinearity, and are obtained following the same strategy.
 
\end{abstract}


\section{Introduction}

In  this paper we investigate the following generalized dispersionless Kadomtsev - Petviashvili equation 
\beq\label{dKPmn}
\ba{l}
\left( u_t +u^mu_x\right)_x+\triangle_{\bot}u=0 , \ \ u=u(x,\vec{y}, t)\in\RR, , \\
\ \  \\
\triangle_{\bot} := \sum_{i=1}^{n-1} {\partial^2_{y_i}}, \ \ \ \vec{y}=(y_1, \cdots, y_{n-1})\in\RR^{n-1}, \ \ 
x,t\in\RR ,
\ea
\eeq
in $(n+1)$ dimensions, where $(m+1)$ is the degree of the nonlinearity, and $m,n\in\NN^+$. Hereafter, we refer to (\ref{dKPmn}) as the $dKP(m,n)$ equation. Equations (\ref{dKPmn}) contain, as particular cases, the integrable (through the method of characteristics) Riemann equation \cite{whitham}
\beq\label{Riemann_equ}
u_t +u^mu_x=0
\eeq
for $n=1$, the integrable dispersionless Kadomtsev - Petviashvili (dKP) equation \cite{LRT,timman,zobolot,kadom} 
\beq\label{dKP}
(u_t +uu_x)_x+u_{yy}=0
\eeq
for $(m,n)=(1,2)$, the nonintegrable Khokhlov - Zobolotskaya (KZ) equation \cite{zobolot} 
\beq
\label{KZ}
(u_t +uu_x)_x+u_{{y_1}{y_1}}+u_{{y_2}{y_2}}=0
\eeq 
for $(m,n)=(1,3)$, and the nonintegrable modified dKP (mdKP) equation \cite{WAS} 
\beq\label{mdKP}
(u_t +u^2u_x)_x+u_{yy}=0
\eeq
for $(m,n)=(2,2)$. 

Equation (\ref{dKPmn}) describes weakly nonlinear and quasi one-dimensional waves, with negligeable dissipation and dispersion, if the linear approximation of the original theory is given by the $(n+1)$ - dimensional wave equation, at least in some limiting cases like the long wave approximation \cite{MS_dkpn}. In equation (\ref{dKPmn}) there are two competing terms: the nonlinear term $u^m u_x$, responsible for the steepening of the profile, and the $\triangle_{\bot}u$ term, describing diffraction in the transversal $(n-1)$ dimensional hyperplane (reminiscence, through the multiscale expansion leading to (\ref{dKPmn}), of the diffraction described by the wave operator); therefore diffraction increases as $n$ increases \cite{MS_dkpn}. The degree $(m+1)$ of the nonlinearity originates from expanding the nonlinear terms of the original PDEs in power series, when the first $m$ powers of the expansion are absent due, usually, to some symmetry of the problem. In this sense, the degree of universality of (\ref{dKPmn}) decreases as the degree of nonlinearity grows (see \S 2). $dKP(1,n),~n=2,3$ are relevant in acoustics \cite{LRT,timman,zobolot,MS_dkpn}, and also in plasma physics \cite{kadom,MS_dkpn,thesis} in the long wave approximation; $dKP(1,2)$ in hydrodynamics, again in the long wave approximation \cite{abl segur,abl clarks}; $dKP(1,3)$ and $dKP(2,3)$ in nonlinear optics \cite{thesis} and $dKP(2,2)$ in the study of sound waves in antiferromagnets \cite{TF}.

We remark that equation (\ref{dKPmn}) arises as the  $x$-dispersionless limit of the natural generalization in $n+1$ dimensions and with nonlinearity of degree $m+1$: 
\beq\label{KPmn}
\left( u_t +u_{xxx}+u^mu_x\right)_x+\triangle_{\bot}u=0 , \ \ u=u(x,\vec{y}, t)\in\RR. 
\eeq
of the $2+1$ dimensional Ka\-dom\-tsev-Pet\-via\-shvi\-li (KP) equation \cite{kadom, novikov, abl segur, abl clarks}.

Apart from the Riemann equation (\ref{Riemann_equ}), integrable by the method of characteristics, in the family  of equations (\ref{dKPmn}) only the dKP equation (\ref{dKP}) is integrable through a novel Inverse Scattering Transform (IST) for integrable dispersionless PDEs \cite{MS_arxiv,MS_IST_heav,MS_IST_dKP}, recently made rigorous in \cite{GSW} on the example of the Pavlov equation \cite{pavlov}. This IST allows one to show, in particular, that solutions $u(x,y,t)$ of dKP depend on $x$ through the combination $x-2ut$ \cite{MS_RH}; i.e., these solutions can be written in the characteristic form \cite{MS_RH,MS_finite}
\beq\label{charact_dKP}
u=F(\zeta,y,t), \ \ \zeta=x-2 F(\zeta,y,t)t,
\eeq
in analogy with the case of the Riemann equation (\ref{Riemann_equ}), for which the dependence of the solution $u(x,t)$ on $x$ is through the combination $x-u^mt$. For this reason, the IST for dKP can be viewed as a significant generalization of the method of characteristics. The formulation (\ref{charact_dKP}) has allowed one to study in an analytically explicit way the interesting features of the gradient catastrophe of two dimensional waves at finite time \cite{MS_RH,MS_finite} and in the longtime regime \cite{MS_RH} in terms of the initial data. 

The other examples of $dKP(m,n)$ equations are not integrable; therefore the possibility to investigate a generic wave breaking through equations like (\ref{charact_dKP}) and the precise form that these equations should take are, in our opinion, challenging open 
problems; in addition, blow up of the solutions is expected for sufficiently large $m$ \cite{WAS} to complicate the picture. In the recent paper \cite{DGK}, f.i., the formal dependence $x-u^m t$, motivated by the $y$ independent limit, was used  to study the generic breaking features of $dKP(m,2)$ and its dispersive shock formation.

In our paper we give some light on the problem of finding a convenient characteristic form of the type (\ref{charact_dKP}) for the study of wave breaking of $dKP(m,n)$ solutions, i) from the construction of a family of exact solutions of $dKP(m,n)$ exhibiting wave breaking,  and ii) from the construction  of the longtime behavior of solutions of the Cauchy problem of $dKP(m,n)$, for small initial data. 

Indeed, after showing in \S 2 the universality of (\ref{mdKP}) starting from a family of nonlinear wave equations, in \S 3 we use the invariance of the $dKP(m,n)$ equations under motions on the paraboloid, to construct a family of exact solutions involving an arbitrary function of one variable, and describing waves constant on their paraboloidal wave front, breaking simultaneously in all points of it, and developing, after breaking, either multivalued overturning profiles or single valued discontinuous shocks. Then we use in \S 4 such solutions to build a uniform approximation of the solution of the Cauchy problem, in the longtime regime and for small and localized initial data, showing that such a small and localized data evolving according to the $dKP(m,n)$ equation break, in the longtime regime, iff $m(n-1)\le 2$. In \S 5 we study the analytic aspects of such a wave breaking, given explicitely in terms of the initial data, providing, in particular, a description of the overturning profile and of the development of a discontinuous shock immediately after breaking; we concentrate, in particular, on the mdKP (\ref{mdKP}) case and on its comparison with the already known dKP (\ref{dKP}) case. The results of this paper, contained in the 2012 master thesis of one of the authors (FS) \cite{thesis}, generalize analogous ones for the $dKP(1,n)$ equation in \cite{MS_dkpn}, and are obtained following the same strategy.

We end this introduction considering the following consequence of these results: for the family of exact solutions of $dKP(m,n)$ and for the longtime asymptotics of the solutions of the small data Cauchy problem, the dependence of the solutions $u(x,\vec y,t)$ on $x$ is, f.i., through the combinations $x-c^{-1}_{m,n}u^m t$, if $c_{m,n}> 0$, and $x-u^m t\log t$, if $c_{m,n}= 0$, where
\beq\label{defcmn}
c_{m,n}=1-\frac{m(n-1)}{2}.
\eeq 
Therefore this dependence involves in a simple way the degree $m+1$ of the nonlinearity and the dimensionality $n+1$ of the problem through the coefficient $c_{m,n}$, and it is consistent with the case of the Riemann equation (\ref{Riemann_equ}), for which $c_{m,1}=1$, and with the dKP equation (\ref{dKP}), for which $c_{1,2}=1/2$ (confirming the apparently mysterious factor $2$ in (\ref{charact_dKP})); it is not consistent, instead, with the formal dependence $x-u^m t$ used in \cite{DGK}.

In analogy with the dKP case, the results of our paper seem to suggest the following conjecture (admittedly a weak conjecture, due to the absence, in the generic case, of an IST generalizing the method of characteristics). \\
{\it Conjecture}. The solutions of the $dKP(m,n)$ equation around breaking, for $c_{m,n}\ge 0$, are described by the following characteristic formulae
\beq
\ba{l}
u=G(\zeta,y,t), \\
\zeta=\left\{
\ba{ll}
x-{c^{-1}_{m,n}}G^m(\zeta,\vec y,t) t, & \mbox{if } c_{m,n}> 0, \\
x-G^m(\zeta,\vec y,t) t \log t, & \mbox{if } c_{m,n}= 0.
\ea
\right.
\ea
\eeq

This paper is dedicated to the memory of S. V. Manakov.
\section{Universality and applicability of $dKP(m,n)$}

The universality (and therefore the applicability) of (\ref{dKPmn}), through a multiscale expansion, is well described by the following simple model: the family of nonlinear wave equations
\begin{equation}\label{mini modl}
(f(w))_{TT} = \triangle w, \qquad \triangle = \sum_{i=1}^{n} \partial^2_{X_i}, \qquad w=w(\vec X, T). 
\end{equation}
For {\bf small amplitudes}: $w \to \epsilon w$, $0<\epsilon \ll 1$, we substitute in (\ref{mini modl}) the nonlinear term by its Taylor expansion
\begin{equation}\label{Taylor}
f(\eps w) = f(0) + f'(0) \epsilon w + \frac{1}{2} f''(0) \epsilon^{2} w^{2} + \cdots .
\end{equation}
obtaining, at $O(\eps)$, the wave equation 
\begin{equation}
w_{TT}=c^2 \triangle w, \qquad  c = 1/ \sqrt{f'(0)}, \label{sk}
\end{equation}
where $f'(0)$ is assumed to be positive. 

If the waves are {\bf quasi one-dimensional} and we choose $X_1$ as the direction of propagation, the wave lengths in the trasversal directions are small: $\vec{k}_{\bot} = \epsilon^{\alpha} \vec{\kappa}_{\bot}$, where $\vec{k}_{\bot}$ is the transversal wave vector and $\alpha>0$ has to be fixed. Then the dispersion relation becomes \cite{MS_dkpn} 
\beq
\omega  = c \sqrt{k_1^2 + \vec{k}_{\bot}^2} 
= ck_1 \sqrt{1+ \epsilon^{2\alpha} \frac{ {\vec{\kappa}_{\bot}}^2  }{k_1^2}} \simeq c k_1 \left( 1+ \epsilon^{2\alpha} \frac{ \vec{\kappa}_{\bot}^2}{2 k_1^2} \right),
\eeq
and the phase of a monochromatic wave reads
\beq
\vec{k} \cdot \vec{X} - \omega T = k_1 \left( X_1 - c T \right)  + \epsilon^{\alpha} \vec{\kappa}_{\bot} \cdot \vec{X}_{\bot} - \frac{c \epsilon^{2\alpha}}{2} 
\frac{ {  {\vec{\kappa}}_{\bot}  }^2  }{k_1^2}  T ,
\eeq
motivating the introduction of the new variables
\begin{equation}
\left \{ \begin{array}{rl}
x \  = & X_1 - c T, \\
\vec{y} \  = & \epsilon^{\alpha} \vec{X}_{\bot},~~y_i=X_{i+1},~i=1,\dots,n-1, \\
t \  = & \epsilon^{2\alpha} \frac{c}{2}T .
\end{array} \right.  \label{fb}
\end{equation}
Rewriting (\ref{mini modl}) in the new variables and imposing $\alpha=1/2$ to get the maximal balance, one obtains, at $O(\eps^2)$, the $dKP(1,n)$ equation
\begin{equation}
\left( u_t +uu_x\right)_x+\triangle_{\bot}u=0,
\end{equation}
where $u=-\frac{c^2}{2}f''(0)w$, and $\triangle_{\bot}$ is the tranversal Laplacian given in (\ref{dKPmn}b).

If the term $f''(0)$ vanishes, the maximal balance must involve the cubic term; if also $f'''(0)=0$, the maximal balance must involve the quartic term, and so on. In the very special case in which  $f^{(l)}(0)=0$, $l=2,\ldots m$, and $f^{(m+1)}(0)\ne 0$, then the maximal balance must involve the term of order $m+1$, and is achieved for $\alpha=m/2$ at the first nontrivial order $O(\epsilon^{m+ 1})$, obtaining the $dKP(m,n)$ equation 
(\ref{dKPmn}), 
with $u=C^{1/m} w$ and  
\begin{equation}
C : = - \frac{c^2 f^{(m+1)}(0)}{m!}, \label{gd}
\end{equation}
if $m$ is odd, and
\begin{eqnarray}
\left( u_t - \mathrm{sgn}(f^{(m+1)}(0)) \ u^m u_x \right)_x + \triangle_{\bot} u =0 \label{gc} ,
\end{eqnarray}
with $u=|C|^{1/m} w$, if $m$ is even. 

The above considerations explain well why equations (\ref{dKPmn}) are less and less universal and applicable, as $m$ increases. 

\section{Exact solutions of $dKP(m,n)$}
It was observed in \cite{MS_dkpn} that $dKP(1,n)$ is invariant under the following Lie symmetry group of transformations
\begin{equation}\label{Lie group}
\left \{ \begin{array}{rl}
\tilde{x} \ = & x + \sum_{i=1}^{n-1} {(\delta_i y_i - \delta_i^2 t)},  \\
\tilde{y}_j \  = &y_j - 2 \delta_j t,  \qquad j= 1, \cdots , n-1  \\
\tilde{t} \  = & t 
\end{array} \right. ,
\end{equation}
where $\delta_j,~j=1,\dots,n-1,$ are the arbitrary parameters of the group, leaving invariant the paraboloid
\begin{equation}\label{parabola}
\xi = x + \frac{1}{4t} \sum_{i=1}^{n-1} y_i^2. 
\end{equation}
Such a symmetry was used to construct a family of exact and implicit solutions of $dKP(1,n)$ exhibiting wave breaking and playing a relevant role 
in the longtime regime of the small data Cauchy problem for $dKP(1,n)$ \cite{MS_dkpn}. 

Since $dKP(m,n)$ is also invariant under the transformation (\ref{Lie group}), following the same strategy as in \cite{MS_dkpn}, such a symmetry will be used in this section to construct a family of exact and implicit solutions of $dKP(m,n)$. In this paper we add, to the exact solutions exhibiting a gradient catastrophe and multivaluedness after breaking, also the exact weak solutions of $dKP(m,n)$ developing, after breaking, single valued discontinuous shocks of dissipative nature.

Therefore we look for solutions of (\ref{dKPmn}) in the form
\begin{eqnarray}\label{reduction}
u =v(\xi, t), \ \ \  \xi =x + \frac{1}{4t} \sum_{i=1}^{n-1} y_i^2,
\end{eqnarray}
reducing (\ref{dKPmn}) to the $(1+1)$ dimensional PDE 
\begin{equation}\label{equ:v}
v_t + v^mv_{\xi} + \frac{n-1}{2t} v=0 .
\end{equation}
The following change of variables
\begin{equation}\label{changevar1}
v(\xi,t) = t^{-\frac{n-1}{2}} q (\xi, \tau),
\end{equation}
where
\begin{equation}\label{deftau1}
\tau(t) = \left\{ \begin{array}{ll}
\frac{1}{c_{m,n}} \  t^{c_{m,n}}+ \alpha,  & \mbox{if } c_{m,n} \neq 0\\
\ln t +\beta , & \mbox{if } c_{m,n} =0 
\end{array} \right. , 
\end{equation}
$\alpha,\beta$ are real constants and the coefficient $c_{m,n}$ is defined in (\ref{defcmn}),   
transforms (\ref{equ:v}) into the Riemann equation (\ref{Riemann_equ}) in the variables $(\xi,\tau)$
\beq\label{Riemann}
q_{\tau} + q^mq_{\xi}=0.
\eeq
We recall \cite{whitham} that (\ref{Riemann}) has the general implicit solution
\beq\label{gen sol multiv}
q=A(\zeta), \ \ \zeta=\xi-A^m(\zeta)\tau \ \ \ \Leftrightarrow \ \ \ q = A(\xi- q^m \tau),
\eeq
where $A$ is an arbitrary differentiable function of one argument. If, in particular, $A$ describes a localized positive hump, the solution breaks first at $\tau=\tau_b$, on the characteristic $\zeta_b$, where 
\beq\label{def1_taub}
\tau_b=\mbox{min }_{\zeta}\left(-\frac{1}{mA^{m-1}(\zeta)A'(\zeta)}\right)=-\frac{1}{mA^{m-1}(\zeta_b)A'(\zeta_b)}>0.
\eeq
For $\tau>\tau_b$, the solution becomes multivalued and not acceptable in many physical contexts. Alternatively, for $\tau>\tau_b$, the regular multivalued solution can be replaced by a weak solution, a single valued discontinuous shock of dissipative nature, whose wave front discontinuity $s(\tau)$ is described by equations \cite{whitham}
\beq\label{shock_def}
\ba{l}
\frac{ds}{d\tau}=\frac{1}{m+1}\frac{A^{m+1}({\zeta}_1)-A^{m+1}({\zeta}_2)}{A({\zeta}_1)-A({\zeta}_2)}, \\
\ \ \\
s={\zeta}_{1}+A^m(\zeta_{1})\tau={\zeta}_{2}+A^m(\zeta_{2})\tau , 

\ea
\eeq 
with the initial conditions 
\beq\label{shock_initial}
s(\tau_b)=\xi_b, \ \ \zeta_1(\tau_b)=\zeta_2(\tau_b)=\zeta_b.
\eeq
Behind and ahead the shock, the solution is given by
\beq\label{shock_q_leftright}
q=\left\{
\ba{ll}
q_{2}=A(\zeta_2), & \mbox{if } \xi< s(\tau), \\
q_{1}=A(\zeta_1), & \mbox{if } \xi> s(\tau),
\ea
\right.
\eeq 
where $q_2$ and $q_1$ are respectively the maximum and the minimum among the (three, in the case of a single hump) branches of the implicit equations (\ref{gen sol multiv}).

We remark that $\tau(t)$ in (\ref{deftau1}) is a monotonically increasing function of $t$; to have it positive, we choose the constants $\alpha,\beta$ and the $t$-intervals as follows:  
\begin{equation}\label{deftau2}
\tau(t) = \left\{ \begin{array}{ll}
\frac{1}{c_{m,n}} \  t^{c_{m,n}}, \  t>0,  & \mbox{if } c_{m,n} > 0, \\
\ln t , \ t>1, & \mbox{if } c_{m,n} =0, \\
\frac{1}{|c_{m,n}|}\left(t^{-|c_{m,n}|}_0-t^{-|c_{m,n}|} \right), \ t>t_0, & \mbox{if } c_{m,n} < 0, 
\end{array} \right.  
\end{equation}

Recalling (\ref{reduction}), (\ref{changevar1}), and (\ref{deftau2}), the exact solutions of $dKP(m,n)$ in the original variables corresponding to (\ref{gen sol multiv}) read
\beq\label{exactsolmultiv}
u = \left\{ \begin{array}{ll}
t^{-\frac{n-1}{2}} A \left( x + \frac{1}{4t} \sum_{i=1}^{n-1} y_i^2 - \frac{1}{c_{m,n}} u^m t\right), \ t>0, & \mbox{if }  c_{m,n}>0 \\
t^{-\frac{n-1}{2}} A \left( x + \frac{1}{4t} \sum_{i=1}^{n-1} y_i^2 - u^m t \ln t \right), \ t>1, & \mbox{if }  c_{m,n}=0 \\
t^{-\frac{n-1}{2}} A \left( x + \frac{1}{4t} \sum_{i=1}^{n-1} y_i^2 - \frac{1}{|c_{m,n}|} u^m t
\left(\left(\frac{t}{t_0}\right)^{|c_{m,n}|}-1\right)\right), \ t>t_0,  & \mbox{if }  c_{m,n}<0,
\end{array} \right. 
\eeq 
where $c_{m,n}$ is defined in (\ref{defcmn}). 

If, again, $A$ is a localized hump, (\ref{exactsolmultiv}) describes a wave constant on the paraboloidal wave front (\ref{parabola}) and simultaneously breaking on it. More precisely, the solution breaks at time $t_b$ defined by
\begin{equation}\label{tb_cmn positive}
t_b = \left\{ \begin{array}{ll}
(c_{m,n}\tau_b)^{1/c_{m,n}},  & \mbox{if } c_{m,n} > 0, \\
e^{\tau_b} , & \mbox{if } c_{m,n} =0 , 
\end{array} \right.  
\end{equation}
on the paraboloid
\beq\label{parab_break}
x + \frac{1}{4t} \sum_{i=1}^{n-1} y_i^2=\zeta_b+A^m(\zeta_b)\tau_b,
\eeq
where $\tau_b,\zeta_b$ are defined in (\ref{def1_taub}) (see Figures 1).

\vskip 10pt
\noindent
\begin{center}
\mbox{\epsfxsize=4.3cm \epsffile{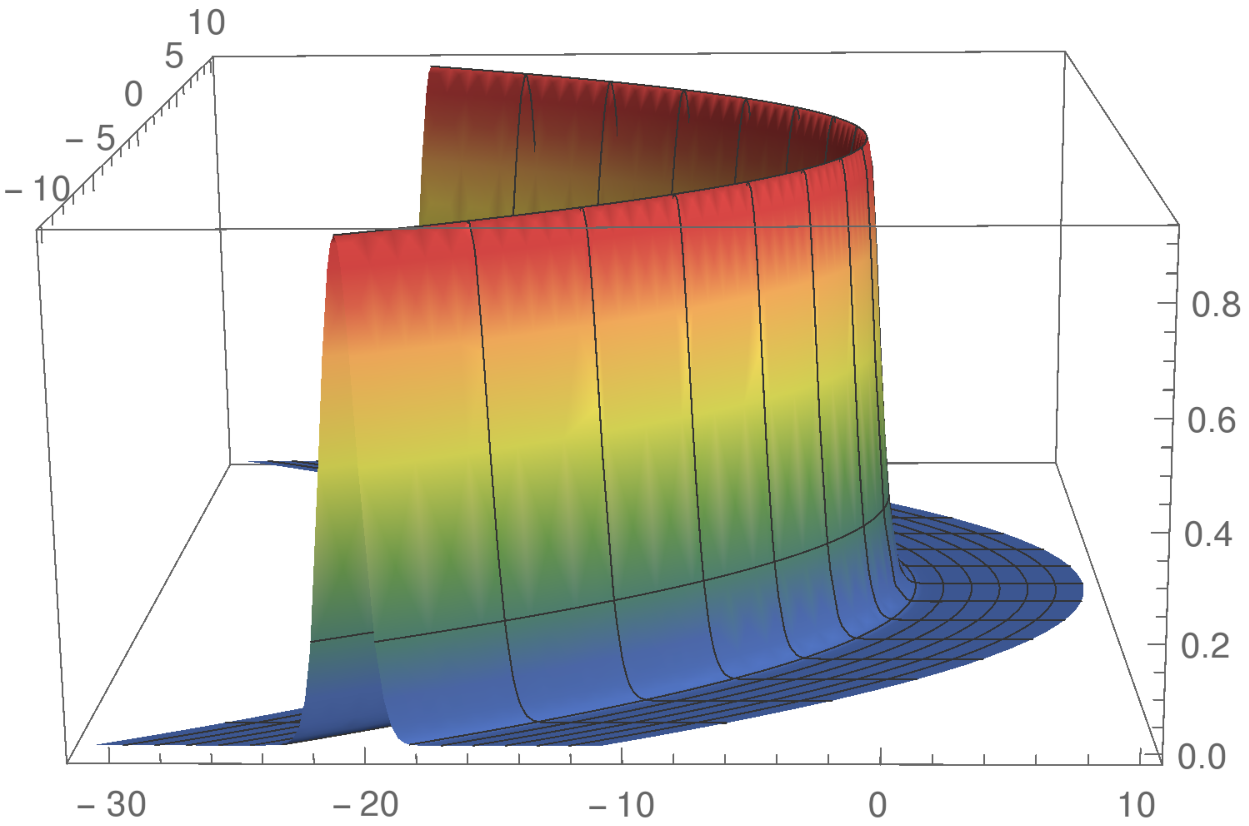}}
\mbox{\epsfxsize=4.3cm \epsffile{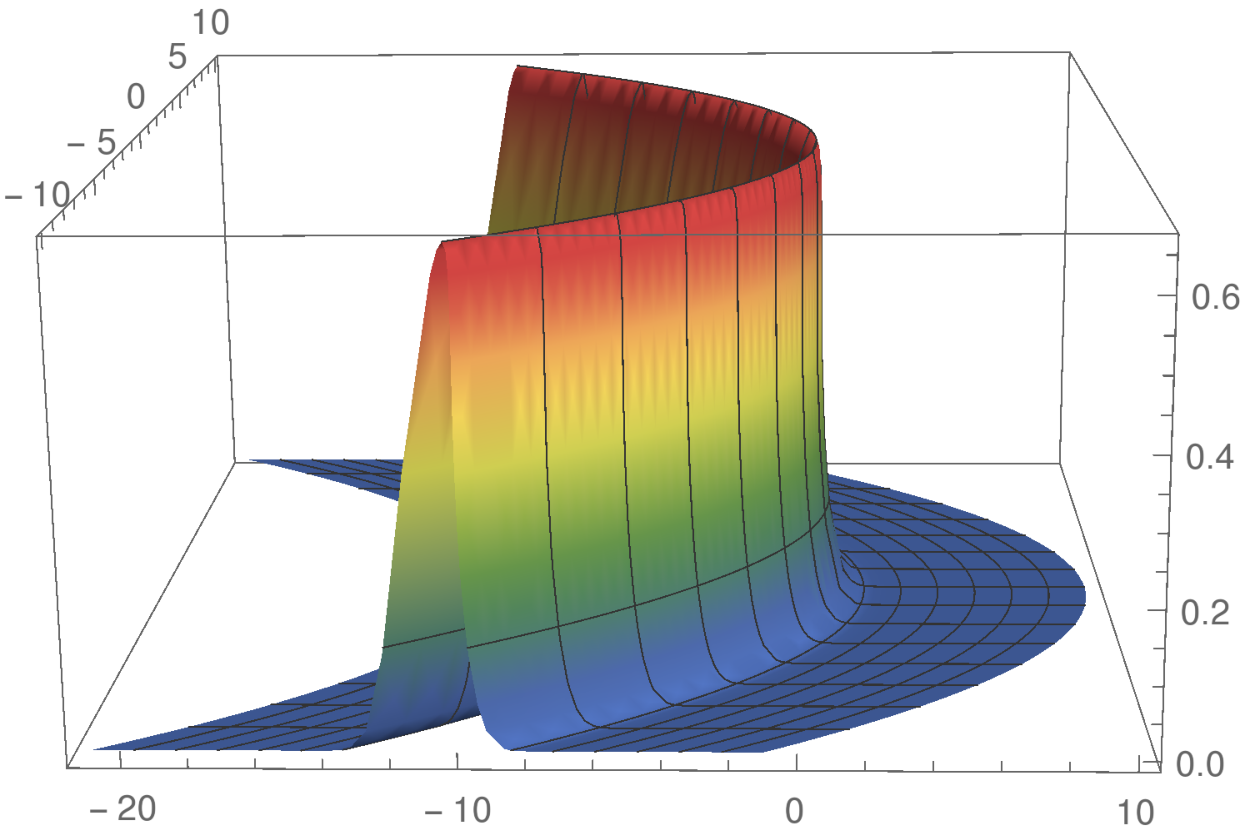}}
\mbox{\epsfxsize=4.3cm \epsffile{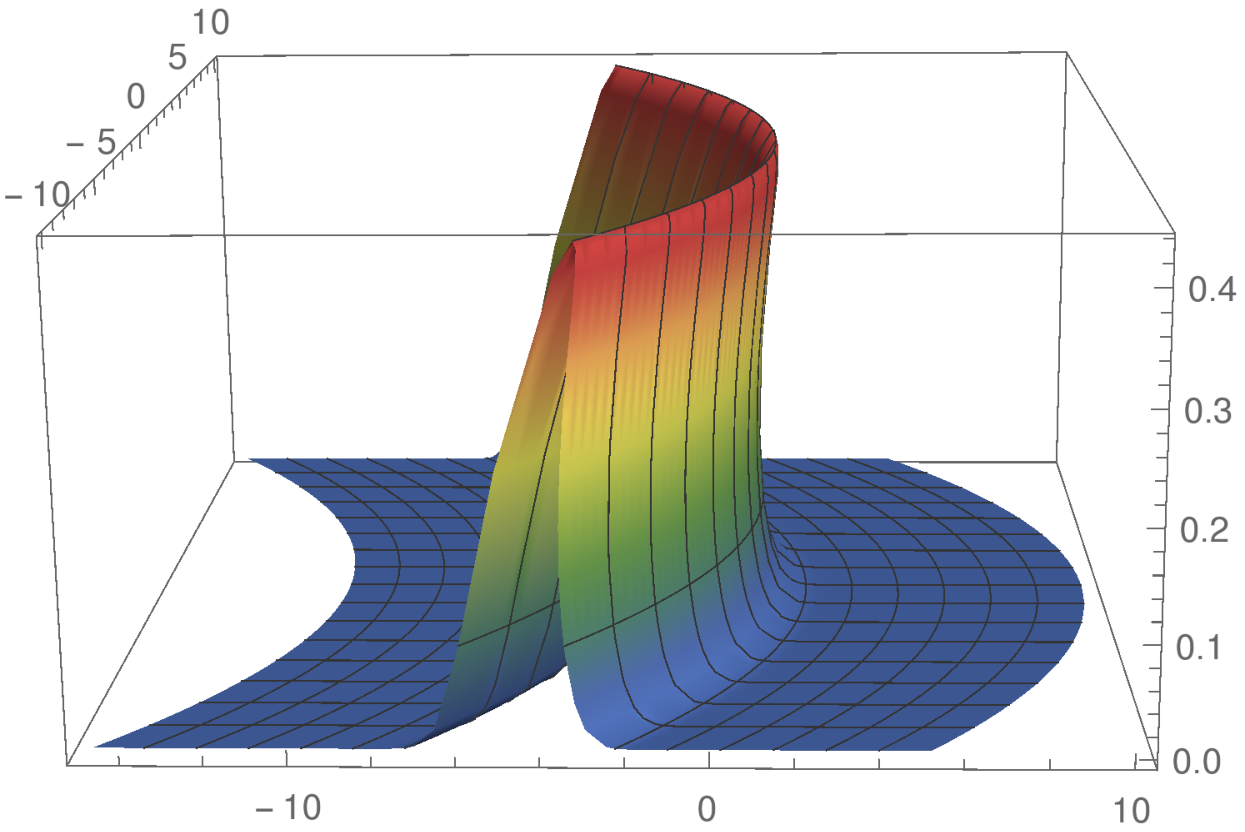}} \\
\end{center}

\noindent
Figures 1. Plotting of the exact solution of the mdKP equation (\ref{mdKP}) ($c_{2,2}=0$), before, at and after breaking. Here we have chosen $A(x)=e^{-x^2}$; consequently: $\zeta_b=1/2,~\tau_b=\sqrt{e}/2,~t_b=e^{\sqrt{e}/2}$.
\vskip 10pt
If $c_{m,n} < 0$, $\tau(t)\to \tau(\infty)\equiv |c_{m,n}|^{-1}t^{-|c_{m,n}|}_0$ as $t\to\infty$, and the wave freezes asymptotically.  
If $t^{-|c_{m,n}|}_0>\tau_b|c_{m,n}|$, then $\tau(\infty)>\tau_b$,  and the wave breaks before freezing at 
\beq\label{tb_cmn negative}
t_b=\left(t^{-|c_{m,n}|}_0-\tau_b|c_{m,n}|\right)^{-1/|c_{m,n}|} ,
\eeq 
on the paraboloid (\ref{parab_break}). If, instead, $t^{-|c_{m,n}|}_0<\tau_b|c_{m,n}|$, then $\tau(\infty)<\tau_b$ and no breaking takes place before the wave freezes.

If $n=1,~\forall m$, (\ref{exactsolmultiv}) reduces to the general solution $u=A(x-u^m t)$ of the Riemann equation (\ref{Riemann_equ}); if $m=1,~\forall n$, (\ref{exactsolmultiv}) reduces to the class of particular solutions of the dKP equation with quadratic nonlinearity and arbitrary dimensions ($dKP(1,n)$) constructed in \cite{MS_dkpn}. It is interesting to remark that this type of exact solutions was first derived for the integrable dKP equation (\ref{dKP}) using its nonlinear Riemnann - Hilbert inverse problem \cite{MS_solv}.   

After breaking: $t>t_b$, the regular multivalued solutions (\ref{exactsolmultiv}) can be replaced by the discontinuous single valued solutions of $dKP(m,n)$ obtained using (\ref{reduction}), (\ref{changevar1}), and (\ref{deftau2}) in (\ref{shock_def}), (\ref{shock_initial}), (\ref{shock_q_leftright}). More precisely, for $t>t_b$, where $t_b$ is defined in (\ref{tb_cmn positive}) and (\ref{tb_cmn negative}), the $dKP(m,n)$ solution described by (\ref{exactsolmultiv}) for $t<t_b$, develops a discontinuous shock on the parabola
\beq
\ba{l}
x+\frac{y^2}{4t}=S(t), \ \ \ 
S(t)\equiv s(\tau(t)), 
\ea
\eeq
where
$s(\tau)$ is characterized by equations (\ref{shock_def}), (\ref{shock_initial}), and $\tau(t)$ is defined in (\ref{deftau2}). Behind and ahead the shock, the solution is given by
\beq\label{shock_u_leftright}
u=\left\{
\ba{ll}
u_{2}(x,\vec y,t), & \mbox{if } x+\frac{y^4}{4t}< S(t), \\
u_{1}(x,\vec y,t), & \mbox{if } x+\frac{y^4}{4t}> S(t),
\ea
\right.
\eeq 
where $u_2$ and $u_1$ are respectively the maximum and the minimum among the (three, in the case of a localized bump) branches $u_{1,2}=t^{-(n-1)/2}A(\zeta_{1,2})$ of the implicit equations (\ref{exactsolmultiv}).

 We end this section remarking that, as already mentioned in \S 1, in the above class of exact solutions of (\ref{dKPmn}) the dependences of $u$ on $x$ through the combinations $x-c^{-1}_{m,n}u^m t$ (if $c_{m,n}> 0$), $x-u^m t\log t$ (if $c_{m,n}= 0$), and 
$x-|c_{m,n}|^{-1} u^m t[(t/t_0)^{|c_{m,n}|}-1]$, if $c_{m,n}<0$, are not consistent with the formal dependence $x-u^m t$ recently postulated to study the breaking features of $dKP(m,2)$ in \cite{DGK}.
 
\section{The Cauchy problem for small and localized initial data and longtime wave breaking}
\label{cauchy}
In analogy with \cite{MS_dkpn}, now we use the exact solutions of the previous section to construct the longtime behaviour of solutions of the Cauchy problem of (\ref{dKPmn}) for small and localized initial data, showing in particular that {\bf small and localized initial data evolving according to (\ref{dKPmn}) break, in the longtime regime, only if $m(n-1)\le 2$ ($c_{m,n}\ge 0$)}. 

For small and localized initial data of the type 
\begin{equation}\label{u0}
u(x, \vec{y}, 0) = \epsilon u_0(x, \vec{y}), \ \ 0 < \epsilon \ll 1 
\end{equation}
evolving according to equation (\ref{dKPmn}), the behavior of the solution may be approximated by the solution of the linearized $dKP(m,n)$ equation until nonlinearity becomes relevant. 

For $O(\epsilon)$ initial data, equations (\ref{Riemann}),(\ref{def1_taub}) imply that the breaking time is  
\beq\label{breakregime1}
\tau_b = O(\epsilon^{-m}) \gg 1 ;
\eeq
so, the nonlinear regime for $dKP(m,n)$ is characterized by the condition 
\begin{equation}\label{breakregime2}
t = O\left(\tau^{-1}(\epsilon^{-m})\right) = \left \{ \begin{array}{ll}
O\left(\epsilon^{-\frac{m}{c_{m,n}}}\right), & c_{m,n}> 0\\
O\left({e}^{\epsilon^{-m}}\right), & c_{m,n} =0 
\end{array} \right. 
\end{equation}
where $\tau^{-1}(\cdot)$ is the inverse of (\ref{deftau1}), while, for $c_{m,n}< 0$, large values of $t$ are not compatible with (\ref{breakregime1}), and a nonlinear regime never occurs. 

So, for  $t \ll \tau^{-1}(\epsilon^{-m})$, the solution of the  Cauchy problem is approximated by the solution of the linearized $dKP(m,n)$ equation, while for $t \sim \tau^{-1}(\epsilon^{-m})$, a matching will be made between the linearized solution and  the exact solution  (\ref{exactsolmultiv}). 

In the linear regime, when $0< t \ll O(\tau^{-1}(\epsilon^{-m}))$, the $dKP(m,n)$ solution is well described by that of the linearized $dKP(m,n)$, i.e.: 
\begin{equation}
u(x, \vec{y}, t) \sim \frac{\epsilon}{(2 \pi )^n} \int_{\mathbb{R}^n} \hat{u}_0 (k_1, \vec{k}_{\bot}) \exp \left[ \imath (k_1 x + \vec{k}_{\bot} \cdot \vec{y} - \frac{k_{\bot}^2}{k_1} t) \right] \mathrm{d}k_1 \mathrm{d}\vec{k}_{\bot} \label{n}
\end{equation}
where
\begin{displaymath}
\hat{u}_0 (k_1, \vec{k}_{\bot}) = \int_{\mathbb{R}^n} u_0 (x, \vec{y}) \exp \left[- \imath (k_1 x + \vec{k}_{\bot} \cdot \vec{y}) \right] \mathrm{d}x \mathrm{d}\vec{y} .
\end{displaymath}
In the longtime regime 
\beq\label{linear regime}
1\ll t \ll O(\tau^{-1}(\epsilon^{-m})),
\eeq  
the multiple integral can be evaluated by the stationary phase method, which gives \cite{MS_dkpn} 
\begin{equation}\label{longtimelinear}
u (x, \vec{y}, t) \sim t^{-\frac{n-1}{2}} \epsilon G \left( x + \frac{1}{4t} \sum_{i = 1}^{n-1} y_i^2, \frac{\vec{y}}{2t} \right),
\end{equation}
where
\begin{equation}\label{defG}
G(\xi, \vec{\eta}) := \frac{1}{2^n \pi^{\frac{n+1}{2}}} \int_{\mathbb{R}} \mathrm{d}k_1 |k_1|^{\frac{n-1}{2}} \hat{u}_0 \left( k_1, \vec{\eta}k_1 \right) \exp \left[ \imath k_1 \xi -\imath(n-1) \frac{\pi}{4} \mathrm{sgn}(k_1) \right] . 
\end{equation}
in the space-time region
\begin{equation}
(x-\xi)/t, \ y_i/t =O(1), \quad i=1,\ldots, n-1 ,
\end{equation}
on the paraboloid (\ref{parabola}), and decays faster outside it. Therefore, {\bf small and localized initial data set, in the longtime regime (\ref{linear regime}) governed by the linear theory, on the paraboloid (\ref{parabola})}.

When nonlinearity becomes relevant, i.e., when  $t= O(\tau^{-1}(\epsilon^{-m}))$, the asymptotic solution of $dKP(m,n)$ is built matching  the longtime solution (\ref{longtimelinear}),(\ref{defG}) of the linearized Cauchy problem with the exact solution (\ref{exactsolmultiv}). If, f.i., $c_{m,n}>0$, the first argument $\xi$ of function $G(\xi,\eta)$ defined in (\ref{defG}) is replaced by $\xi-{c^{-1}_{m,n}}u^m t$, or, equivalently, the arbitrary function $A(\zeta)$ in (\ref{gen sol multiv}) acquires the dependence on the second argument $\vec\eta=\vec y /2t$, and is identified with $G$. As a result of this matching, in the nonlinear regime (\ref{breakregime1}), (\ref{breakregime2}) the solution reads 
\begin{equation}\label{uas}
\ba{l}
u(x, \vec{y}, t)\simeq u^{as}_{m,n}(x, \vec{y}, t), \\
u^{as}_{m,n}(x, \vec{y}, t) \equiv  \left \{ \begin{array}{ll}
t^{-\frac{n-1}{2}} \epsilon G \left( x + \frac{1}{4t} \sum_{i = 1}^{n-1} {y_i^2} - \frac{1}{c_{m,n}} u^m t, \frac{\vec{y}}{2 t} \right), &  c_{m,n} > 0 , \\
t^{-\frac{n-1}{2}} \epsilon G \left( x + \frac{1}{4t} \sum_{i = 1}^{n-1} y_i^2 -  u^m t \ln t , \frac{\vec{y}}{2 t} \right), & c_{m,n} = 0 ,
\end{array} \right.
\ea
\end{equation}
with $u^mt=O(1)$ for $c_{m,n} > 0$, and $u^mt\log t=O(1)$ for $c_{m,n} = 0$, becoming infinitesimal in the linear regime, as it has to be.

For $c_{m,n} < 0$, the solution is described by its linear form (\ref{longtimelinear}); therefore wave breaking takes place for the particular values of $(m,n)$ described by the condition
\beq\label{breakmn}
c_{m,n} \ge 0 \ \ \ \Leftrightarrow \ \ \ m(n-1)\le 2 ;
\eeq
i.e., for
\begin{equation} 
\begin{tabular}{|l|l|l|}
\hline
$n=1$ & $n=2$ & $n=3$\\
$\forall m$ & $m=1,2$ & $m=1$\\
\hline
\end{tabular} \label{x}
\end{equation}
For $n=1,~\forall m$, $u^{as}_{m,n}$ reduces to the general solution of the Riemann equation (\ref{Riemann_equ}) \cite{whitham}. The cases $(m,n)=(2,1)$ and $(m,n)=(3,1)$, the dKP and the KZ equations, have been already investigated in \cite{MS_RH} and \cite{MS_dkpn} respectively; therefore, in the following section we mainly focus on the case $(m,n)=(2,2)$, corresponding to the nonintegrable modified dKP equation (\ref{mdKP}) and to its asymptotic solution 
\begin{equation}\label{asympt_mdKP}
u\simeq u^{as}_{2,2}(x,y,t) = \frac{1}{\sqrt{t}} \epsilon G \left( x + \frac{y^2}{4t}  -  u^2 t \ln t , \frac{y}{2 t} \right) ,
\end{equation}
comparing the results with the dKP case $(m,n)=(1,2)$ investigated in \cite{MS_RH}, with its asymptotic solution
\beq\label{asympt_dKP}
u\simeq u^{as}_{1,2}(x,y,t) = \frac{1}{\sqrt{t}} \epsilon G \left( x + \frac{y^2}{4t}  - 2  u t, \frac{y}{2 t} \right) 
\eeq

\vskip 10pt
\noindent  
{\bf Remark 1}. The estimate of the first correction to the asymptotics (\ref{uas}) reads 
\beq\label{error}
u=u^{as}_{n,m}(x,\vec{y},t)(1+ O(t^{-1})). 
\eeq
Indeed, if $c_{m,n}>0$, one has  
\beq\label{error1}
u\sim u^{as}_{n,m}(x, \vec{y}, t)
+\frac{\eps^{1+\frac{m}{c_{m,n}}}}{t^{\frac{n-1}{2}}} 
H\left(x + \frac{1}{4t} \sum_{i = 1}^{n-1} {y^2_i},\frac{\vec{y}}{2 t},\check\tau \right),
\eeq
where $\check\tau =\eps^m \tau$ and $H(\xi,\vec\eta,\check\tau)$ is expressed in terms of $G(\zeta,\vec\eta)$ ($\zeta$ is defined in the second of equations (\ref{charact})), through the PDE 
\beq
\left(\frac{H_{\check\tau}+G^m H_{\xi}}{1+m G^{m-1}G_{\zeta}\check\tau} \right)_{\xi}+
\frac{1}{4(c_{m,n}\check\tau)^{1+c^{-1}_{m,n}}}\sum\limits_{j=1}^{n-1}
\left(\frac{G_{\eta_j}}{1+m G^{m-1}G_{\zeta}\check\tau} \right)_{\eta_j}=0.
\eeq
Since, from (\ref{breakregime2}), $\eps^{\frac{m}{c_{m,n}}}=O(t^{-1})$, equation (\ref{error1}) yields (\ref{error}). Similarly, if $c_{m,n}=0$,
\beq\label{error2}
u\sim u^{as}_{n,m}(x, \vec{y}, t)
+\frac{\eps}{t^{\frac{n+1}{2}}} 
H\left(x + \frac{1}{4t} \sum_{i = 1}^{n-1} {y^2_i},\frac{\vec{y}}{2 t},\check\tau \right),
\eeq
where $H(\xi,\vec\eta,\check\tau)$ now solves the PDE
\beq
\left(\frac{H}{1+m G^{m-1}G_{\zeta}\check\tau} \right)_{\xi}=
\frac{1}{4}\sum\limits_{j=1}^{n-1}
\left(\frac{G_{\eta_j}}{1+m G^{m-1}G_{\zeta}\check\tau} \right)_{\eta_j}.
\eeq
Then equation (\ref{error2}) yields equation (\ref{error}) too.  
\vskip 10pt
\noindent 
{\bf Remark 2}. As is was observed in \cite{MS_dkpn}, for a gaussian initial condition 
\begin{equation}\label{gaussian}
u_0(x,y) = d~\exp \left[-\frac{x^2 + {\vec y}^2}{4}\right],
\end{equation}
where $d$ is constant, the asymptotic solution can be written in terms of special functions, obtaining for $G(\xi, \vec{\eta})$:
\beq\label{G_gaussian}
\ba{l}
G(\xi,\vec{\eta})=\\ 
\frac{d}{\sqrt{\pi}}\frac{1}{\left(1+{\vec\eta}^2\right)^{\frac{n+1}{4}}}
\Big[\cos\left(\frac{\pi}{4}(n-1)\right) \Gamma\left(\frac{n+1}{4}\right) \ _1F_1 \left(\frac{n+1}{4},\frac{1}{2},-\frac{Y^2}{4}\right) \\
+\sin\left(\frac{\pi}{4}(n-1)\right)\Gamma\left(\frac{n+3}{4}\right)Y \ _1F_1 \left(\frac{n+3}{4},\frac{3}{2},-\frac{Y^2}{4}\right)\Big] 
\ea
\eeq 
where
\beq
Y=\frac{\xi}{\sqrt{1+{\vec\eta}^2}},
\eeq
$\Gamma$ is the Euler  gamma function and $\,_1F_1$ is the Kummer  confluent hypergeometric function \cite{abramowitz, bateman}. For $n=2$ and $d=\sqrt{2\pi}$, it yields, respectively:
\beq\label{G_gaussian_n=2}
\ba{l}
G(\xi,{\eta})=\\ 
\frac{1}{\left(1+{\eta}^2\right)^{\frac{3}{4}}}\left[
\Gamma\left(\frac{3}{4}\right) \ _1F_1 \left(\frac{3}{4},\frac{1}{2},-\frac{Y^2}{4}\right)+
Y\Gamma\left(\frac{5}{4}\right) \ _1F_1 \left(\frac{5}{4},\frac{3}{2},-\frac{Y^2}{4}\right)\right]. \\
\ea
\eeq 

\vskip 5pt
\noindent
{\bf Remark 3}. The fact that small data break only when the inequality (\ref{breakmn}) is satisfied was expected from physical considerations. Indeed, as already mentioned in the introduction, in equation (\ref{dKPmn}) there are two competing terms: the nonlinear term $u^m u_x$, responsible for the steepening of the profile, and the $\triangle_{\bot}u$ term, describing diffraction in the transversal $(n-1)$ dimensional hyperplane (reminiscence, through the multiscale expansion leading to (\ref{dKPmn}), of the wave operator); therefore diffraction increases as $(n-1)$ increases \cite{MS_dkpn}. For small initial data, the $u^m u_x$ term is initially smaller and smaller increasing $m$, and the solution evolves in a linear way for a longer and longer time, diffracting transversally through the $(n-1)$ diffraction channels. So, diffraction, increasing with $(n-1)$, acts on a very long time, increasing with $m$, before the nonlinear regime could become relevant, and if $(n-1)$ and/or $m$ are sufficiently large, one expects that the solution would be diffracted away almost completely, before reaching the nonlinear regime, and will not break. \\
We have, in particular, that breaking takes place in the following longtime regimes (see (\ref{breakregime2})): \\
$m=n=1$ (the Riemann equation with quadratic nonlinearity): $t_b=O(\eps^{-1})$; \\
$m=1,~n=2$ (the dKP equation): $t_b=O(\eps^{-2})$; \\
$m=1,~n=3$ (the KZ equation): $t_b=O(e^{\eps^{-1}})$; \\
$m=2,~n=1$ (the Riemann equation with cubic nonlinearity): $t_b=O(\eps^{-2})$; \\ 
$m=2,~n=2$ (the mdKP equation): $O(t_b=e^{\eps^{-2}})$. 

It is interesting to remark that, as it was observed in \cite{MS_dkpn}, if $m=1$ (the physically more relevant case), wave breaking takes place only for $n\le 3$; i.e., in physical space!         

We end this section remarking that also in the longtime behaviour of the solutions of the small data Cauchy problem for equation (\ref{dKPmn}), the dependence of the solution $u$ on $x$, as in the case of the exact solutions (\ref{exactsolmultiv}), is not consistent with the formal ansatz recently made in \cite{DGK}.


\section{Analytic aspects of the wave breaking}
\label{break}
In this section we use the longtime behavior of solutions of the small data Cauchy problem for $dKP(m,n)$ to show explicitly the analytic aspects of the solution in the neighbourhood of the breaking time in terms of the initial data, represented by function G defined in (\ref{defG}), as it was already done in \cite{MS_RH,MS_dkpn}.

Choosing $c_{m,n}\ge 0$, we rewrite equation (\ref{uas}) in the characteristic form
\beq\label{charact}
\ba{l}
q\sim \epsilon G(\zeta, \vec\eta), \\
\xi = \eps^m F(\zeta, \vec\eta;m) \tau + \zeta , \ \ F(\zeta, \vec\eta;m)\equiv G^m(\zeta, \vec\eta),
\ea
\eeq
where  
\beq\label{transf_uq}
q=u t^{\frac{n-1}{2}}, \qquad \xi = x + \frac{1}{4t} \sum\limits_{j=1}^{n-1}y^2_j, \qquad \vec \eta = \frac{\vec y}{2t}
\eeq
and
\begin{equation}\label{deftau3}
\tau(t) = \left\{ \begin{array}{ll}
\frac{1}{c_{m,n}} \  t^{c_{m,n}},  & \mbox{if } c_{m,n} > 0, \\
\ln t , & \mbox{if } c_{m,n} =0. 
\end{array} \right.  
\end{equation}

One solves the second of equations (\ref{charact}) 
with respect to the parameter $\zeta$, obtaining $\zeta(\xi,\vec\eta,\tau),$ 
and replaces it into the first, to obtain the solution  
$q\sim \eps G(\zeta(\xi,\vec\eta,\tau),\vec\eta)$. The inversion of the second of equations (\ref{charact}) 
is possible iff its $\zeta$-derivative is different from zero. Therefore the singularity manifold (SM) of 
the solution  is the $n$ - dimensional manifold characterized by the equation 
\beq\label{SM}
{\cal S}(\zeta,\vec\eta,\tau)\equiv 1+\eps^m F_{\zeta}(\zeta,\vec\eta,m)\tau=0~~~
\Rightarrow~~~\tau=-\frac{1}{\eps^m F_{\zeta}(\zeta,\vec\eta,m)}.
\eeq
Since 
\beq
\label{grad q}
\ba{l}
\nabla_{(\xi,\vec\eta)}q=\frac{\eps \nabla_{(\zeta,\vec\eta)}G(\zeta,\vec\eta)}{1+\eps^m F_{\zeta}(\zeta,\vec\eta,m)\tau},
\ea
\eeq
the slope of the localized wave becomes infinity (the so-called gradient catastrophe) on the SM, 
and the $n$-dimensional wave ``breaks'' except on the direction represented by the vector field 
$\hat V=\sum\limits_{i=1}^{n-1}\partial_{\eta_i}+m \eps^m G^{m-1}(\sum\limits_{i=1}^{n-1}G_{\eta_i})\partial_{\xi}$ \cite{MS_dkpn}, \cite{MS_finite}, on which:
\beq\label{transversal}
\hat V q=\eps \sum\limits_{i=1}^{n-1}G_{\eta_i}.
\eeq

Since the KZ case $(m,n)=(1,3)$ has been studied in detail in \cite{MS_dkpn}, in the following we mainly concentrate on the mdKP case $(m,n)=(2,2)$, and we compare it with the dKP case $(m,n)=(1,2)$ investigated in \cite{MS_RH}. 

From  (\ref{SM}), the first breaking time $\tau_b$ and the corresponding characteristic parameters $\vec{\zeta}_b =(\zeta_b,\eta_b)$  are defined by the global minimum of a function of the $2$ variables $(\zeta,\eta)$:
\begin{equation}\label{def_taub}
\tau_b= - \frac{1}{\epsilon^m F_\zeta(\zeta_b, \eta_b;m)}:= \inf \left(- \frac{1}{\epsilon^m F_\zeta(\zeta, \eta;m)}\right)>0,
\end{equation}  
and it is characterized, together with the condition $F_{\zeta}(\vec {\zeta}_b;m)<0$, by the following equations:  
\beq\label{break_characteriz}
\ba{ll}
F_{\zeta\zeta}(\vec {\zeta}_b;m) = F_{\zeta\eta}(\vec {\zeta}_b;m) =0 \\
\alpha  \equiv F_{\zeta\zeta \zeta}(\vec {\zeta}_b;m) F_{\zeta\eta \eta}(\vec {\zeta}_b;m) -F^2_{\zeta\zeta \eta}(\vec {\zeta}_b;m)  > 0, \qquad F_{\zeta\zeta \zeta}(\vec {\zeta}_b;m) >0.
\ea
\eeq
The corresponding point at which the first wave breaking 
takes place is, from (\ref{charact}), ${\vec \xi}_b=(\xi_b,\eta_b)\in\RR^2$, where:
\beq\label{xb}
\xi_b= {\zeta}_b + \eps^m F({\vec {\zeta}}_b;m)\tau_b ={\zeta}_b -\frac{F({\vec {\zeta}}_b;m)}{F_{\zeta}({\vec {\zeta}}_b;m)}. 
\eeq
Now we evaluate equations (\ref{charact}) and (\ref{SM}) near breaking, in the regime:
\beq
\label{near_breaking}
\xi=\xi_b+\xi',~~\eta=\eta_b+\eta',~~\tau=\tau_b+\tau',~~\zeta={\zeta}_b+{\zeta'},
\eeq
where $\xi',\eta',\tau',\zeta'$ are small. Using (\ref{def_taub}) - (\ref{xb}), the second of  
equations (\ref{charact}) becomes, at the leading order, the following cubic equation in ${\zeta'}$:
\beq\label{cubic}
{\zeta'}^3+a(\eta'){\zeta'}^2+b(\eta',\tilde\tau){\zeta'}-\gamma X(\xi',\eta',\tilde\tau)=0,
\eeq
where
\beq\label{def_a,b,X}
\ba{l}
a(\eta')=\frac{3F_{\zeta\zeta\eta}}{F_{\zeta\zeta\zeta}}\eta', \\
b(\eta',\tilde\tau)=
\frac{3}{F_{\zeta\zeta\zeta}}\left[2 F_{\zeta}\tilde\tau +F_{\zeta\eta\eta}{\eta'}^2\right], \\
X(\xi',\eta',\tilde\tau)=\xi'-\eps F({\zeta}_b,\eta_b+\eta')\tau'-
\eps \left[F({\zeta}_b,\eta_b+\eta')-F\right]\tau_b  \sim \\
\xi'-\frac{F_{\eta}}{|F_{\zeta}|}\eta'-\frac{F}{|F_{\zeta}|} \tilde\tau-
\frac{F_{\eta\eta}}{2|F_{\zeta}|}{\eta'}^2 -\frac{F_{\eta}}{|F_{\zeta}|}\eta'\tilde\tau -
\frac{F_{\eta\eta\eta}}{6|F_{\zeta}|}{\eta'}^3,~~~~
\gamma=\frac{6|F_{\zeta}|}{F_{\zeta\zeta\zeta}},
\ea
\eeq
and  
\beq\label{def_tildetau}
\tilde\tau \equiv \frac{\tau'}{\tau_b}=\frac{\tau -\tau_b}{\tau_b},
\eeq
corresponding to the maximal balance
\beq\label{max_balance}
|{\zeta'}|,|\eta'|=O(|\tilde\tau |^{1/2}),~~~|X|=O(|\tilde\tau |^{3/2}).
\eeq
In (\ref{def_a,b,X}) and in the rest of this section, $G,~F=G^m$ and  all their partial derivatives whose 
arguments are not indicated are meant to be evaluated at $\vec {\zeta}_b=(\zeta_b,\eta_b)$.

The three roots of the cubic are given by the well-known Cardano-Tartaglia formula:
\beq\label{sol_cubic1}
\ba{l}
{\zeta'}_0\left(\xi',\vec\eta',\tilde\tau\right)=-\frac{a}{3}+(A_+)^\frac{1}{3} + (A_-)^\frac{1}{3} , \\
{\zeta'}_{\pm}\left(\xi',\vec\eta',\tilde\tau\right)=-\frac{a}{3}-\frac{1}{2}\left((A_+)^\frac{1}{3} + 
(A_-)^\frac{1}{3}\right) \pm \frac{\sqrt{3}}{2}i\left((A_+)^\frac{1}{3} - (A_-)^\frac{1}{3}\right),  
\ea
\eeq
where
\beq\label{def_A}
\ba{l}
A_{\pm}=R\pm \sqrt{\Delta}
\ea
\eeq
and the discriminant $\Delta$ reads
\beq
\label{discriminant}
\Delta=R^2+Q^3,
\eeq
with
\beq
\label{def_QR} 
\ba{l}
Q(\eta',\tilde\tau)=\frac{3b-a^2}{9}=-\frac{2 |F_{\zeta}|}{F_{\zeta\zeta\zeta}}\tilde\tau + 
\frac{\alpha}{F^2_{\zeta\zeta\zeta}}{\eta'}^2,       \\
R(\xi',\eta',\tilde\tau)=\frac{\gamma}{2}X(\xi',\eta',\tilde\tau)+\frac{ab}{18}+
\frac{a}{3}Q(\eta',\tilde\tau).
\ea
\eeq
At the same order, function $\cal S$ in (\ref{SM}) becomes
\beq
\label{SM_2}
\ba{l}
{\cal S}(\zeta',\eta',\tilde\tau)=-\tilde\tau+
\frac{1}{2|F_{\zeta}|}\Big[F_{\zeta\zeta\zeta}{\zeta'}^2+2F_{\zeta\zeta\eta}{\zeta'}\eta'+
F_{\zeta\eta\eta}{\eta'}^2\Big]= \\
\frac{F_{\zeta\zeta\zeta}}{2|F_{\zeta}|}\left(Q(\eta',\tilde\tau)+\left(\zeta' +\frac{F_{\zeta\zeta\eta}}{F_{\zeta\zeta\zeta}}\eta' \right)^2 \right).
\ea
\eeq
Known ${\zeta'}$ as function of ($\xi',\eta',\tilde\tau$) from the cubic (\ref{cubic}), 
the solution $q$ and its gradient are then approximated, near breaking, 
by the formulae:
\beq
\label{sol_gradsol}
\ba{l}
q(\xi,\eta,\tau)\sim \eps G({\zeta}_b+{\zeta'},\eta_b+\eta')=\eps \left(G+G_{\zeta}\zeta' +G_{\eta}\eta'+
O(|\tilde\tau |)\right),    \\
\nabla_{(\xi,\eta)}q\sim \eps \frac{\nabla_{(\zeta',\eta')}G({\zeta}_b+{\zeta'},\eta_b+\eta')}
{{\cal S}(\zeta',\eta',\tilde\tau)}.
\ea
\eeq

\vskip 5pt
\noindent
{\bf Before  breaking}
\vskip 5pt

Before breaking: $\tau<\tau_b$ $(\tilde\tau <0)$, the coefficient $Q$ in (\ref{def_QR}) is strictly positive (see the second of (\ref{break_characteriz})); then the discriminant $\Delta=R^2+Q^3$ is also strictly positive  
and only the root ${\zeta'}_0$ is real. Correspondingly, the real solution $q$ is 
single valued and described by Cardano's formula and by (\ref{sol_gradsol}). In addition, function $\cal S$ in (\ref{SM_2}) 
is also strictly positive and $\nabla_{(\xi,\vec\eta)}q$ is finite $\forall ~\xi,\eta$.

To have a more explicit solution than that provided by Cardano's formula, we first restrict the asymptotic region 
to a narrower area, so that the cubic (\ref{cubic}) reduces to the linear equation $b\zeta'=\gamma X$, and $X$ contains only linear terms. This is achieved choosing:
\beq
\ba{l}
\eta'=O({\tilde\tau}^r), \ X=O({\tilde\tau}^{p+1}), \ \frac{1}{2}<p<\frac{p+1}{2}<r<p+1
\ea
\eeq
implying
\beq
\ba{l}
\zeta'=O({\tilde\tau}^{p}),  \\ 
X=\xi'+\frac{F_{\eta}}{F_{\zeta}}\eta'+\frac{F}{F_{\zeta}}\tilde\tau +o({\tilde\tau}^{p+1})= 
\xi'+\frac{G_{\eta}}{G_{\zeta}}\eta'+\frac{G}{m G_{\zeta}}\tilde\tau +o({\tilde\tau}^{p+1}), \\
b\zeta'=\gamma X \ \ \Rightarrow \ \ \zeta'\sim -\frac{1}{\tilde\tau}\left(\xi'+\frac{G_{\eta}}{G_{\zeta}}\eta'+\frac{G}{m G_{\zeta}}\tilde\tau\right)
\ea
\eeq
Then 
\beq
q=\eps G(\zeta_b +\zeta',\eta_b+\eta')\sim  \eps G + \eps G_{\zeta}\zeta'\sim 
\eps\left(1-\frac{1}{m}\right) G - 
\frac{\eps G_{\zeta}}{\tilde\tau}\left(\xi'+\frac{G_{\eta}}{G_{\zeta}}\eta'\right).
\eeq
Since $q\sim\eps G$, and recalling the definitions (\ref{def_taub}),(\ref{def_tildetau}), it follows that $q$ is described by the exact similarity solution 
\beq
q\sim \left(\frac{\xi+(G_{\eta}/G_{\zeta})\eta'}{\tau'} \right)^{\frac{1}{m}}
\eeq
of equation (\ref{Riemann}).

The cubic simplifies also in the asymptotic region in which $\xi', \eta'$ are of the same order while  $\tau' \leq O(|\eta'|)$, reducing to $\zeta'^3 =  \gamma X$, so that the solution $q$ and its gradient are
\begin{eqnarray}\label{soll}
q \sim \epsilon G(\zeta_b + \sqrt[3]{ \gamma X(\xi', \eta', \tilde{\tau})}, \eta_b + \eta') \\
\nabla_{\xi, \eta}q \sim \frac{\sqrt[3]{ \gamma}}{3}  \frac{\epsilon \nabla_{\zeta, \eta} G }{\sqrt[3]{X^2(\xi', \eta', \tilde{\tau})}}. 
\end{eqnarray}
\vskip 5pt
\noindent
{\bf At breaking}
\vskip 5pt

When  $\tau = \tau_b$, the singularity manifold  $\cal S$ (\ref{SM_2}) is positive everywhere in the $(\xi,\eta)$ plane, except at the 
breaking point $(\xi_b,\eta_b)$ , in which it is zero, together with $Q$, $R$ and the discriminant $\Delta$. The solution $q$ is 
well described again by Cardano's formula and by (\ref{sol_gradsol}). To have a more explicit solution than that provided by Cardano's formula, we choose $\xi', \eta'$ of the same order, the cubic (\ref{cubic}) reduces to $\zeta'^3 = \gamma X$ and equations (\ref{soll}) become
\beq\label{break1}
\ba{l}
q \sim \epsilon G(\zeta_b + \sqrt[3]{ \gamma X(\xi', \eta', 0)}, \eta_b + \eta') \\
\nabla_{\xi, \eta}q \sim \frac{\sqrt[3]{ \gamma}}{3}  \frac{\epsilon \nabla_{\xi, \eta} G}{\sqrt[3]{X^2(\xi', \eta', 0)}}
\ea
\eeq

The expression of $X_b(\xi',\vec\eta')\equiv X(\xi',\vec\eta',0)$ is given in (\ref{def_a,b,X}): $X_b(\xi', \eta') \sim \xi' - \frac{F_{\eta}}{|F_{\zeta}|} \eta'$ if $\xi' \propto \eta'$ but $\xi' \neq (F_{\eta}/F_{\zeta}) \eta'$, while $X(\xi', \eta', 0) = - \frac{F_{\eta \eta}}{2 |F_{\zeta}|}  \eta'^2$ if $\xi' = (F_{\eta}/F_{\zeta}) \eta'$.

Equation (\ref{break1}) implies that all the derivatives of $q$ blow up at $\tau=\tau_b$, in 
the breaking point $(\xi_b,\eta_b)$, with the universal law $X^{-2/3}_b$, except the derivative along the transversal line 
$X_b(\xi',\eta',0)=0$, represented by the vector field $\hat V=\partial_{\eta}-{X_b}_{\eta}\partial_{\xi}$ (compare with (\ref{transversal})), for which \cite{MS_dkpn, MS_finite}
\beq
\hat V q |_{(\xi_b,\eta_b)}=\eps G_{\eta}.
\eeq
\vskip 5pt
\noindent
{\bf After breaking I: the overturning profile}
\vskip 5pt

After breaking, the solution becomes three-valued in a compact region of the $(\xi,\eta)$ - plane (see Figures 3), 
and, in general, does not describe any physics; nevertheless a detailed study of the multivalued region is important, in view of a proper regularization of the models. 

If $\tau > \tau_b$ $(\tilde\tau >0)$, in the regime (\ref{max_balance}), the SM equation ${\cal S}=0$:  
\beq\label{ellipse}
\ba{l}
2|F_{\zeta}|\tilde\tau=
F_{\zeta\zeta\zeta}{\zeta'}^2+2F_{\zeta\zeta\eta}{\zeta'}\eta'+F_{\zeta\eta\eta}{\eta'}^2
\ea
\eeq
describes an ellipsoidal paraboloid in the ($\zeta',\eta',\tilde\tau$) space, with minimum at 
the breaking point $(\vec\xi_b,\tilde\tau_b)$.  

One can check that, eliminating $\zeta'$ from equations (\ref{ellipse}) and (\ref{cubic}), one obtains the SM equation in space-time coordinates, coinciding with the equation $\Delta =R^2+Q^3=0$, where $\Delta$ is the discriminant (\ref{discriminant}) of the cubic (\ref{cubic}), and $Q$ and $R$ are defined in (\ref{def_QR}). Such equation can be satisfied only if $Q\le 0$; 
i.e., only if
\beq\label{boundary_eta}
|\eta'|\le \eta'_{cusp}(\tilde\tau),  
\eeq
where
\beq\label{def_etacusp}
\eta'_{cusp}(\tilde\tau)\equiv \sqrt{\frac{2|F_{\zeta}|F_{\zeta\zeta\zeta}}{\alpha}}\sqrt{\tilde\tau},
\eeq
and is given by the following analytic and universal expression
\beq\label{Delta=0}
\ba{l}
\Big\{
\xi'-\frac{F_{\eta}}{|F_{\zeta}|}\eta'-\frac{F}{|F_{\zeta}|} \tilde\tau-
\frac{F_{\eta\eta}}{2|F_{\zeta}|}{\eta'}^2 -\left( \frac{F_{\eta}}{|F_{\zeta}|}+\frac{F_{\zeta\zeta\eta}}{F_{\zeta\zeta\zeta}} \right)  \eta'\tilde\tau 
+\frac{1}{6|F_{\zeta}|F^2_{\zeta\zeta\zeta}}\Big[-F_{\eta\eta\eta}F^2_{\zeta\zeta\zeta} \\+ F_{\zeta\zeta\eta}(3F_{\zeta\zeta\zeta}F_{\zeta\eta\eta}-2F^2_{\zeta\zeta\eta})\Big]{\eta'}^3\Big\}^2=
\frac{\alpha^3}{9 F^2_{\zeta} F^4_{\zeta\zeta\zeta}} 
\left( \left(\eta'_{cusp}(\tilde\tau)\right)^2 - {\eta'}^2 \right)^3.
\ea
\eeq 
Equation (\ref{Delta=0}) describes a closed curve with two cusps in the $(\xi,\eta)$ - plane at the points
\beq\label{cusp}
\vec\xi^{\pm}_{cusp}(\tilde\tau)=(\xi_b,\eta_b)+(\xi^{\pm '}_{cusp}(\tilde\tau),\pm \eta'_{cusp}(\tilde\tau)) ,
\eeq
where
\beq\label{def_xicusps} 
\xi^{\pm '}_{cusp}(\tilde\tau)\sim \pm F_{\eta}\sqrt{\frac{2F_{\zeta\zeta\zeta}}{\alpha |F_{\zeta}| }}\sqrt{\tilde\tau}+\left(\frac{F}{|F_{\zeta}|}+
\frac{F_{\eta\eta}F_{\zeta\zeta\zeta}}{\alpha}\right)\tilde\tau\pm B{\tilde\tau}^{3/2}
\eeq
where $B=O(1)$ (see Figures 6 for dKP and 7 for mdKP), corresponding to the conditions $Q=R=0$, on which the three real 
solutions of the cubic coincide (in $1+1$ dimensions, it is not possible to have three coincident solutions!): 
\beq
\zeta_0=\zeta_{\pm}=-\frac{F_{\zeta\zeta\eta}}{F_{\zeta\zeta\zeta}}\eta', \ \ \Rightarrow \ \ q_0=q_{\pm}=\sim \eps \left(G+(G_{\eta}-G_{\zeta}\frac{F_{\zeta\zeta\eta}}{F_{\zeta\zeta\zeta}})\eta'\right).
\eeq
On the remaining part of the closed curve $\Delta=0$, two of the three real branches coincide. Outside the closed curve, $\Delta>0$ and the real solution $q$ is single valued; inside the closed curve, 
$\Delta <0$ and the real solution $q$ is three valued, with the three branches
\beq\label{3sol}
\ba{l}
q= \epsilon G(\zeta_b +\zeta', \eta_b+\eta')\sim \eps (G+G_{\zeta}\zeta'+G_{\eta}\eta'+O({\tilde\tau})), 
\ea
\eeq
where $\zeta'$ are the three real solutions of the cubic (\ref{cubic}) (see, f.i., Figures 2 for dKP).
\vskip 10pt
\noindent
\begin{center}
\mbox{\epsfxsize=6cm \epsffile{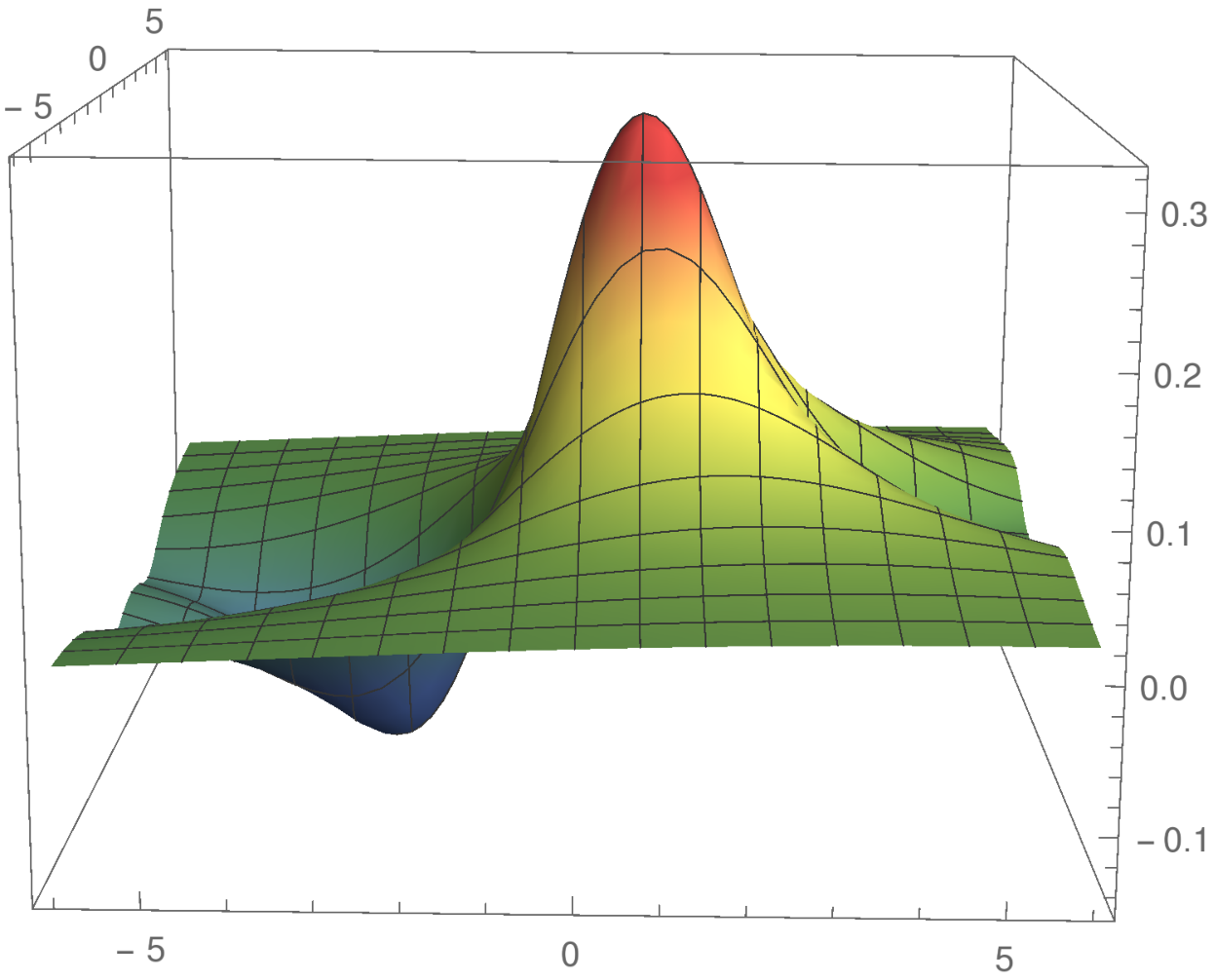}}
\mbox{\epsfxsize=6cm \epsffile{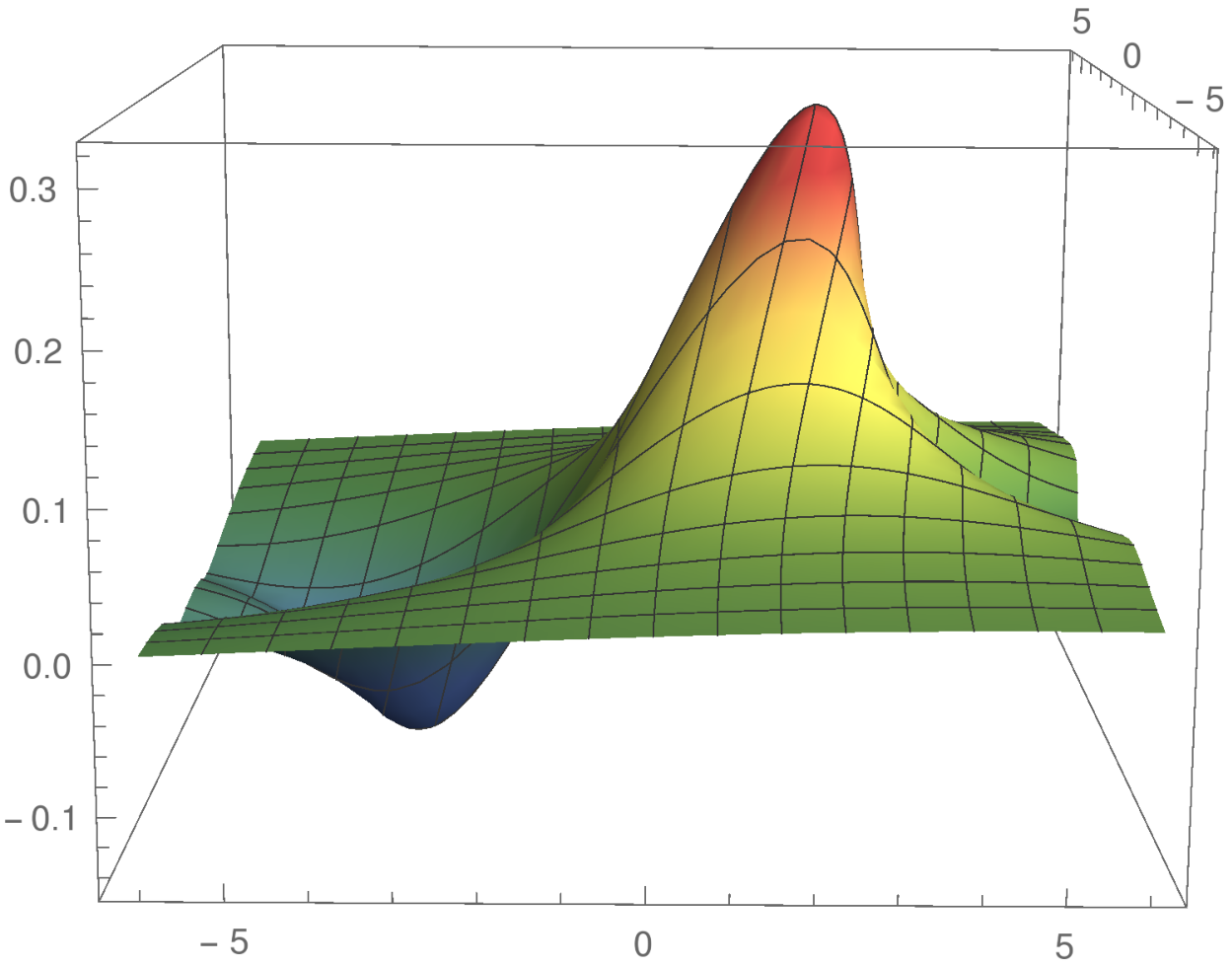}}
\mbox{\epsfxsize=6cm \epsffile{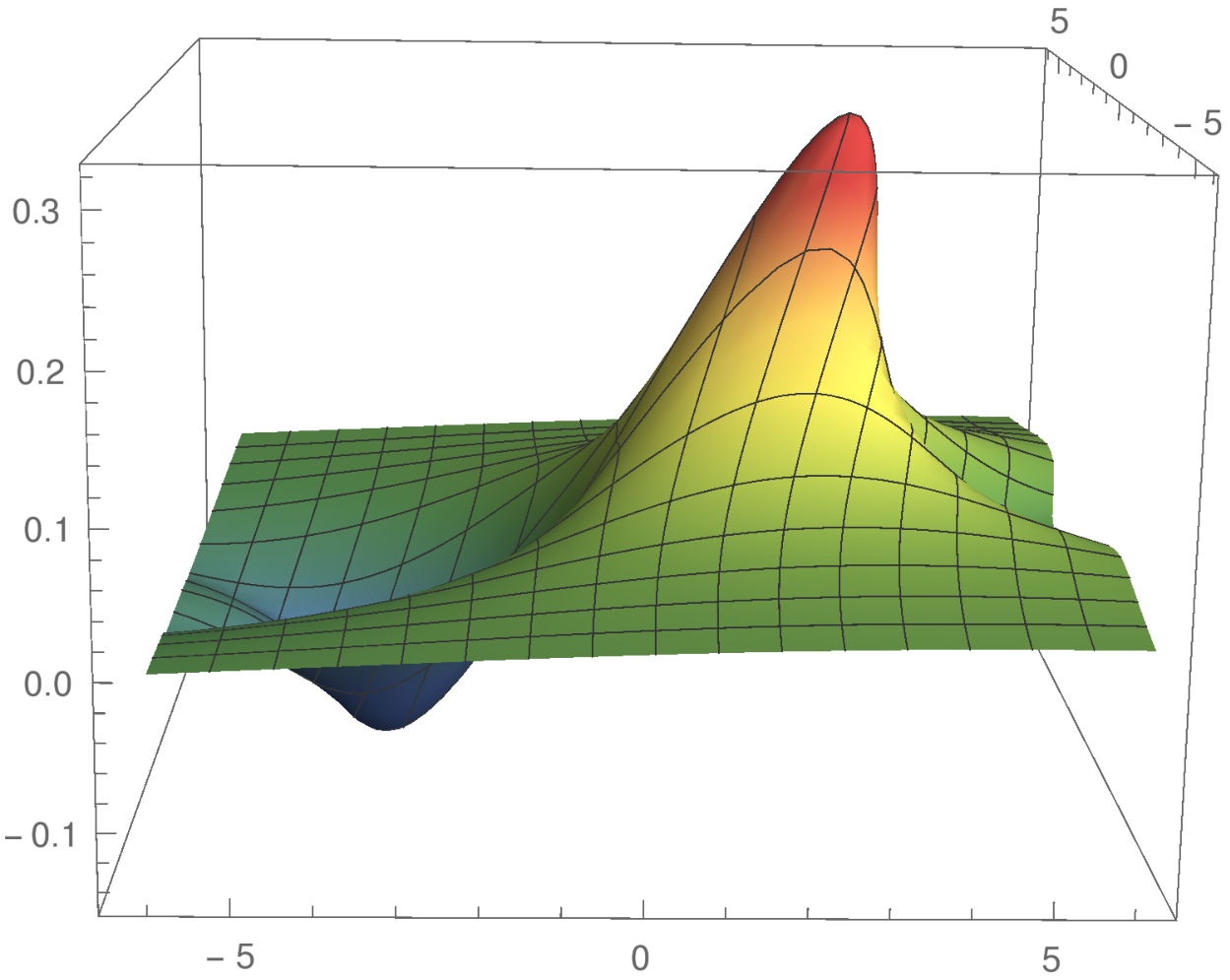}} 
\mbox{\epsfxsize=6cm \epsffile{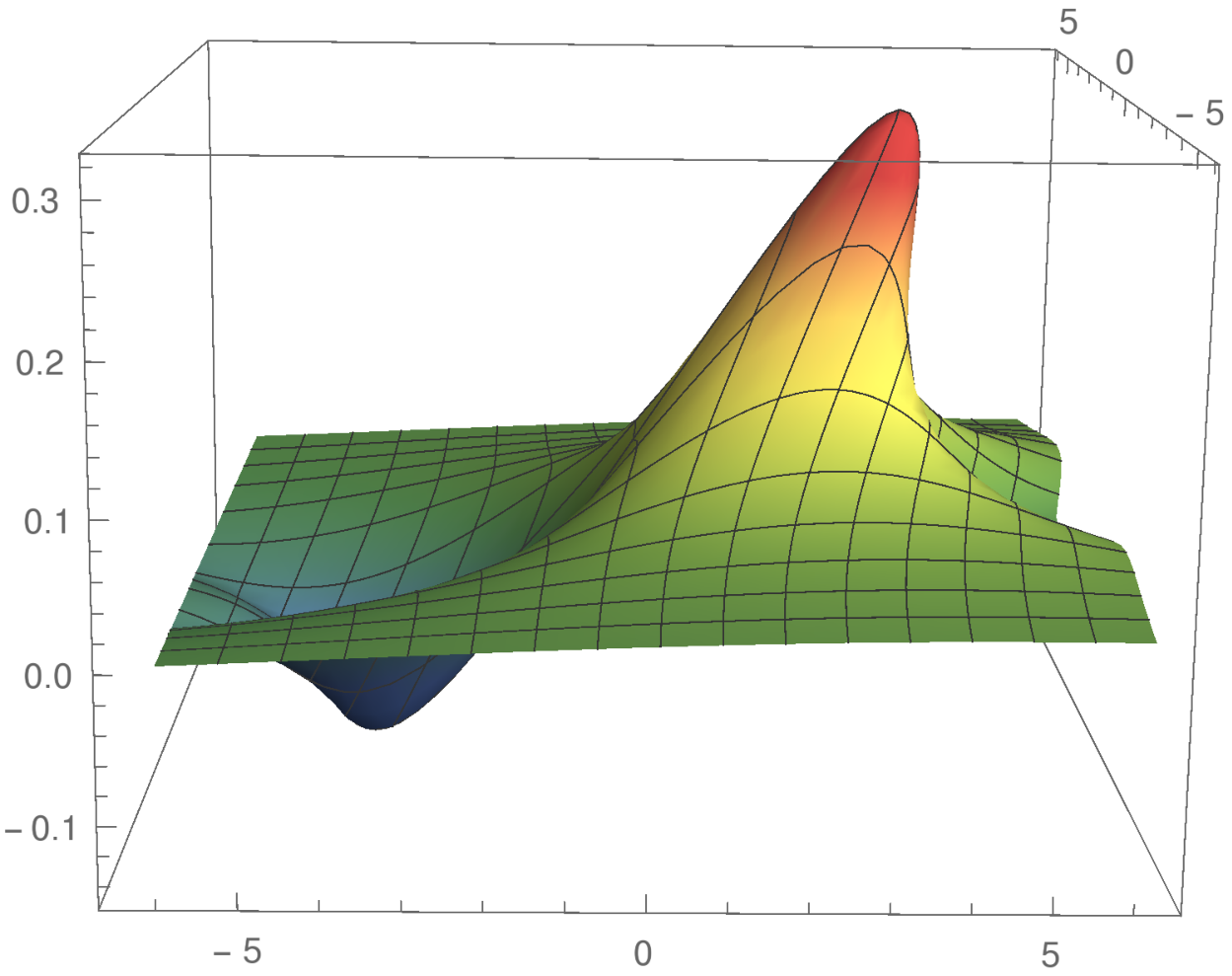}}\\
\end{center}

\noindent
Figures 2. Plotting of the dKP analytic solution in the $(q,\xi,\eta,\tau)$ variables, for $G$ given by (\ref{G_gaussian_n=2}), $\eps=0.2$, at times $\tau =0,~\tau_b-2,~\tau_b,~\tau_b+2.$, where $\tau_b=6.57$. 
\vskip 10pt
Such a three-valued region of the $(\xi,\eta)$ plane is characterized by the conditions
\beq
|\eta' |<\eta'_{cusp}(\tilde\tau), \ \ \xi^-_M(\eta',\tilde\tau)<\xi'<\xi^+_M(\eta',\tilde\tau),
\eeq
where
\beq
\ba{l}
\xi^{\pm}_M(\eta',\tilde\tau)\equiv 
\frac{F_{\eta}}{|F_{\zeta}|}\eta'+\frac{F}{|F_{\zeta}|} \tilde\tau+
\frac{F_{\eta\eta}}{2|F_{\zeta}|}{\eta'}^2 +\left( \frac{F_{\eta}}{|F_{\zeta}|}+\frac{F_{\zeta\zeta\eta}}{F_{\zeta\zeta\zeta}} \right)  \eta'\tilde\tau 
-\frac{1}{6|F_{\zeta}|F^2_{\zeta\zeta\zeta}}\Big[-F_{\eta\eta\eta}F^2_{\zeta\zeta\zeta} \\ +F_{\zeta\zeta\eta}(3F_{\zeta\zeta\zeta}F_{\zeta\eta\eta}-2F^2_{\zeta\zeta\eta})\Big]{\eta'}^3\pm \frac{F_{\zeta\zeta\zeta}}{3 |F_{\zeta}|}|Q|^{\frac{3}{2}}
\ea
\eeq
(see Figures 3 for dKP). 

Since 
\beq\label{longitudilal width}
\Delta\xi \equiv \xi^{+}_M(\eta',\tilde\tau)-\xi^{-}_M(\eta',\tilde\tau)=\frac{2 F_{\zeta\zeta\zeta}}{3 |F_{\zeta}|}|Q(\eta',\tilde\tau)|^{3/2}\le 
\frac{4\sqrt{2}}{3}\sqrt{\frac{|F_{\zeta}|}{F_{\zeta\zeta\zeta}}}{\tilde\tau}^{3/2}, 
\eeq
we infer that the longitudinal width of the multivalued region is $O({\tilde\tau}^{3/2})$; the transversal widths delimited by the two cusps is instead $O({\tilde\tau}^{1/2})$; therefore the multivalued region develops, at $\tau=\tau_b$, 
from the breaking point $(\xi_b,\eta_b)$, with an infinite speed in the tranversal direction, and with zero 
speed in the longitudinal direction. All these results are a common feature to the class of dynamics (\ref{dKPmn}) (see the previously derived results for the dKP equation, for finite times \cite{MS_RH,MS_finite} and in the longtime regime \cite{MS_RH}, and those for the KZ equation in the longtime regime \cite{MS_dkpn}).

\vskip 10pt
\noindent
\begin{center}
\mbox{\epsfxsize=4.3cm \epsffile{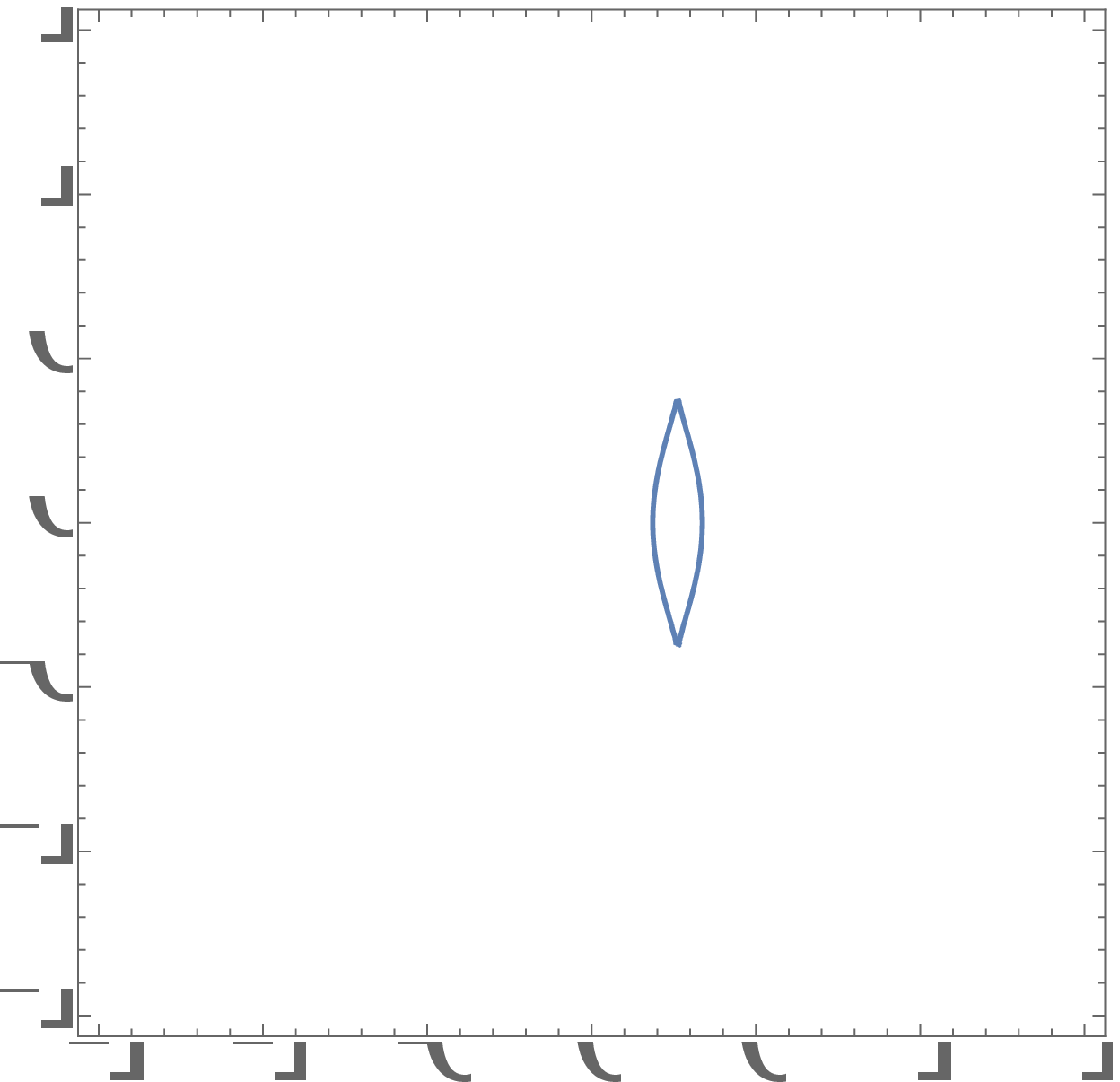}}
\mbox{\epsfxsize=4.3cm \epsffile{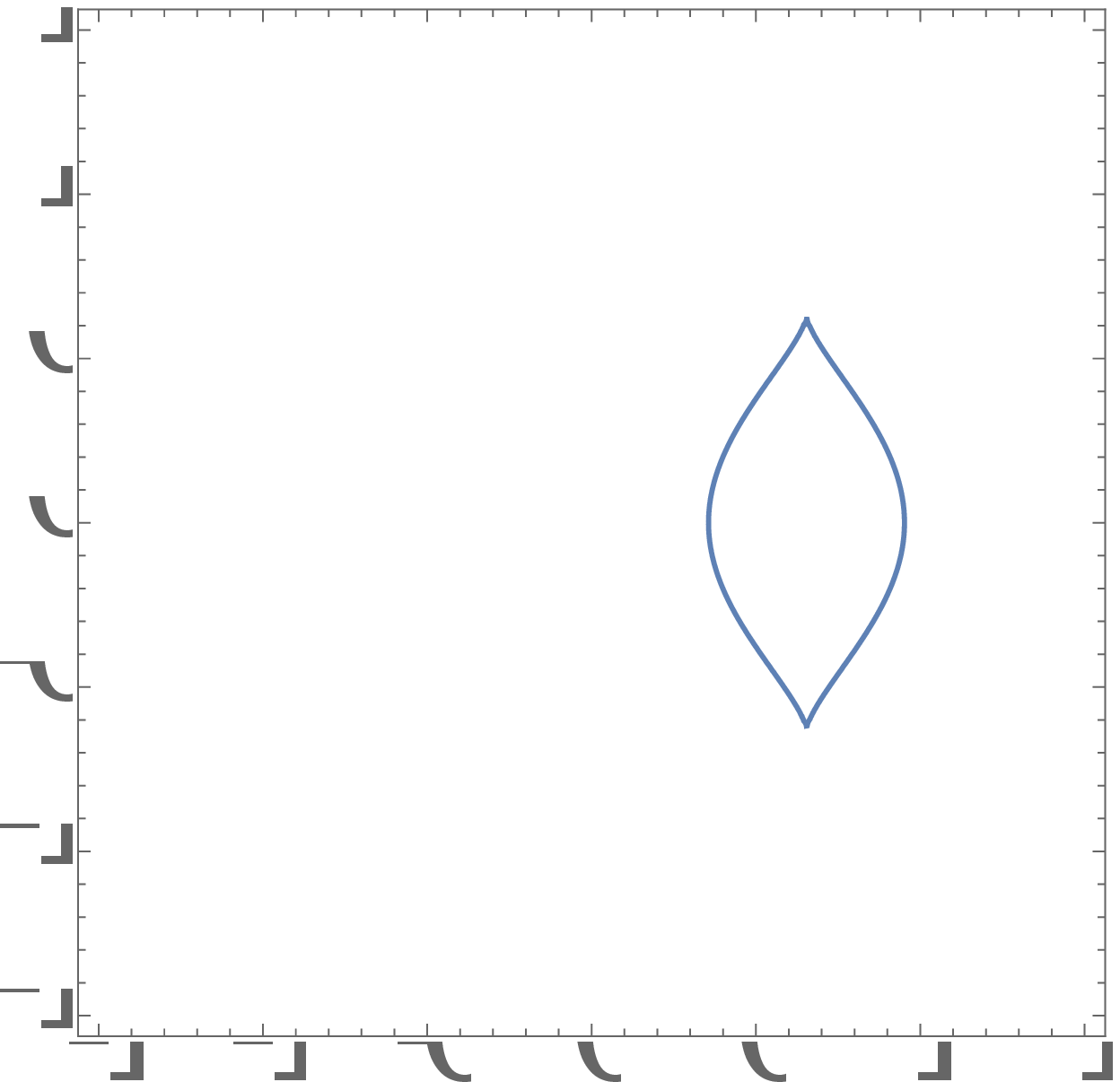}} \\
\end{center}

\noindent
Figures 3. Plotting of the three valued regions of the $(\xi,\eta)$ plane for dKP at times $\tilde\tau =0.2,~0.5$ for $G$ given by (\ref{G_gaussian_n=2}) and $\eps=0.2$. The center of the square frame is the breaking point $(\xi_b,\eta_b)$. Notice that the two cusps have the same $\xi$ coordinate $\xi^{+}_{cusp}(\tilde\tau)\sim \xi^{-}_{cusp}(\tilde\tau)$ since, for solution (\ref{G_gaussian_n=2}), $G_{\eta}=0$ (see (\ref{def_etacusp}),(\ref{def_xicusps})). 

\vskip 5pt
We end this subsection remarking that, as it was already observed in \cite{MS_finite}, since $dKP(m,n)$ is symmetric wrt $y$, if the initial datum is also symmetric wrt $y$, like the Gaussian (\ref{gaussian}), also its evolution will be so: $u(x,y,t)=u(x,-y,t)$. This implies that $\eta_b=y_b=0$, and that all partial derivatives of $F$ and $G$ containing an odd number of $\eta$ derivatives are $0$ at breaking. As a consequence, the transversal direction on which the derivative of the profile does not blow up is the $\eta$ ($y$) direction, and, on it, the derivative is $0$ at breaking (see (\ref{transversal})).

\vskip 5pt
\noindent
{\bf After breaking II: the discontinuous shock}
\vskip 5pt
Using the matching rules of \S 3, the shock formulae (\ref{shock_def}) become
{\setlength\arraycolsep{2pt}
\begin{eqnarray}
\frac{\partial s}{\partial \tau}& =&  \frac{\epsilon^m}{m+1} \frac{G^{m+1}(\zeta_1, \eta)-G^{m+1}(\zeta_2, \eta)}{G(\zeta_1, \eta)-G(\zeta_2, \eta)}, \label{shock_s}\\ 
s &=& \zeta_1 + \epsilon^m G^m(\zeta_1, \eta)\,\tau = \zeta_2 + \epsilon^m G^m(\zeta_2, \eta)\,\tau  \label{shock_zeta}
\end{eqnarray}}
where subscript $2$ is for values behind the shock and subscript $1$ for values ahead. 
Equations (\ref{shock_zeta}) are two implicit equations for $\zeta_1$ and $\zeta_2$: solving these equations yields $\zeta_1(s, \eta, \tau)$, $\zeta_2(s, \eta, \tau)$, and, substituting these expressions in  (\ref{shock_s}), it results in a nonlinear first order differential equation for $s$. This equation and the initial conditions (\ref{shock_initial}) give uniquely $s(\eta, \tau)$ and, consequently, $\check\zeta_{1,2}(\eta, \tau)=\zeta_{1,2}(s(\eta, \tau), \eta, \tau)$. 

We recall that, immediately after breaking and if (\ref{boundary_eta}) holds, equation (\ref{shock_zeta}) reduces to the cubic (\ref{cubic}),(\ref{def_a,b,X}), in which $\xi'$ must be replaced by $s'$, possessing three real roots, where
\begin{equation}\label{def_s'}
s'(\eta', \tilde{\tau}) =s-\xi_b.
\end{equation}
Therefore the values of $\zeta_1$ and $\zeta_2$ are chosen among the three solutions  (\ref{sol_cubic1}) so that  $q_2=\epsilon G(\zeta_b + \zeta'_2, \eta)$ has the maximum value and $q_1=\epsilon G(\zeta_b + \zeta'_1, \eta)$ the  minimum one among the three branches of $q$. Since $G_{\zeta}<0$, it follows that 
\begin{equation}
\zeta_1 = \max\{\zeta'_0, \zeta'_{\pm}\} \qquad \zeta_2 = \min\{\zeta'_0, \zeta'_{\pm}\}. \label{hc}
\end{equation}
To identify $\zeta_{1,2}$ with one of the three roots $\zeta'_0, \zeta'_{\pm}$, let us rewrite them as follows 
{\setlength\arraycolsep{2pt}
\begin{eqnarray}
\zeta'_0 & = & -\frac{a}{3} + 2\sqrt[3]{\rho} \cos \left(\frac{\theta}{3} \right) ,\label{hh}\\
\zeta'_{+} & = & -\frac{a}{3} -\sqrt[3]{\rho} \left( \cos \left(\frac{\theta}{3} \right) +\sqrt{3} \sin\left(\frac{\theta}{3}\right) \right) , \label{hi} \\
\zeta'_{-} & = &   -\frac{a}{3} -\sqrt[3]{\rho} \left(\cos \left(\frac{\theta}{3} \right) - \sqrt{3} \sin\left(\frac{\theta}{3}\right) \right) , \label{hj}
\end{eqnarray}}
where 
\begin{equation}
\rho:= \sqrt{R^2 + |\Delta|}, \qquad \theta:= \arctan\left(\frac{\sqrt{|\Delta|}}{R} \right). \label{hk}
\end{equation}
We observe that, for  $0\leq \theta \leq \pi$, the following inequalities hold 
\begin{displaymath}
-\cos \left(\frac{\theta}{3} \right) - \sqrt{3} \sin\left(\frac{\theta}{3}\right) \leq -\cos \left(\frac{\theta}{3} \right) + \sqrt{3} \sin\left(\frac{\theta}{3}\right) \leq 2 \cos \left(\frac{\theta}{3} \right),
\end{displaymath}
so that
\begin{equation}
\zeta'_+ \leq \zeta'_- \leq \zeta'_0 \Rightarrow q(\zeta_b + \zeta'_0, \eta) \leq q(\zeta_b + \zeta'_-, \eta) \leq q(\zeta_b + \zeta'_+, \eta), \label{ze}
\end{equation}
for $q$ decreasing in  $\zeta$. Hence
\begin{equation}
\zeta'_1(s', \eta', \tau') = \zeta'_0(s', \eta', \tau') , \qquad \zeta'_2 (s', \eta', \tau')= \zeta'_+(s', \eta', \tau') .\label{hd}
\end{equation}
Then (\ref{shock_s}) reduces to the equations
\beq\label{shock_s_local}
\ba{ll}
\frac{ds'}{d\tilde\tau}+\frac{G}{G_{\zeta}}+\frac{G_{\eta}}{G_{\zeta}}\eta'+\frac{1}{2}\left[\zeta'_0(s',\eta',\tilde\tau)+\zeta'_+(s',\eta',\tilde\tau)\right]=0, & \mbox{for dKP}, \\
\frac{ds'}{d\tilde\tau}+\frac{G}{2 G_{\zeta}}+  \frac{G_{\eta}}{G_{\zeta}}\eta'+\frac{1}{2}\left[\zeta'_0(s',\eta',\tilde\tau)+\zeta'_+(s',\eta',\tilde\tau)\right]=0, & \mbox{for mdKP,} \\
s'(0)=\xi_b,
\ea
\eeq
obtained expanding (\ref{shock_s}) near the breaking point and recalling that $\tilde{\tau} \equiv \tau'/\tau_b$.

We remark that, in the differential equations (\ref{shock_s_local}), $s'$ depends on $\eta'$ parametrically. So, the discontinuous (dissipative) shocks of the longtime solutions of the small data Cauchy problems for dKP and mdKP has a simple geometric construction: given $\tilde\tau>0$, for any  fixed $|\eta'|\le \eta_{cusp}$, the overturning profile in the $(\xi,q)$ plane is intercepted by a vertical straight line cutting off equal area lobi, in analogy with the $1+1$ dimensional case.

\vskip 5pt
\noindent
{\bf Going back to the dKP variables}
\vskip 5pt

The above formulae, written in the convenient variables $q,\xi,\eta,\tau$, are essentially the same for both dKP and mdKP; relevant differences appear when we go back to the original variables $u,x,y,t$ inverting equations (\ref{transf_uq}),(\ref{deftau3}): 
\beq
\ba{lllll}
u=\frac{1}{\sqrt{t}}q, & t=\frac{\tau^2}{4}, & y=\frac{\eta\tau^2}{2}, & x=\xi-\frac{\eta^2\tau^2}{4}, & \mbox{for dKP}, \\
u=\frac{1}{\sqrt{t}}q, & t=e^{\tau}, & y=2\eta e^{\tau}, & x=\xi-\eta^2 e^{\tau}, & \mbox{for mdKP},
\ea
\eeq 
and all the above universal formulae are easily transfered to the $dKP(m,n)$ equations. In particular:  
small and localized initial data 
break, i) at $t_b=\tau^2_b/4$ in the point 
$(x_b,y_b)=(\xi_b-\tau^2_b\eta^2_b/4,\eta_b\tau^2_b/2)$ of the parabolic wave front $x+y^2/(4 t_b)=\xi_b$, evolving according to dKP, and ii) at $t_b=e^{\tau_b}$ in the point $(x_b,y_b)=(\xi_b-\eta^2_b e^{\tau_b},2\eta_b e^{\tau_b})$ of the parabolic wave front $x+y^2/(4 t_b)=\xi_b$, evolving according to mdKP. If, for instance, we use the same Gaussian initial condition (\ref{gaussian}) for both dKP and mdKP, with $\eps=0.2$, then $G$ is given by (\ref{G_gaussian_n=2}) in both cases. Consequently, for dKP: 
\beq
\zeta_b=1.96,~\eta_b=0,~\xi_b=3.27,~\tau_b=6.57  
\eeq
and wave breaking takes place at $t_b=10.8$ in the point $(x_b,y_b)=(3.27, 0)$ on the parabola $x+2.31\cdot 10^{-2}y^2=3.27$ (see Figure 4). 
\vskip 10pt
\noindent
\begin{center}
\mbox{\epsfxsize=6cm \epsffile{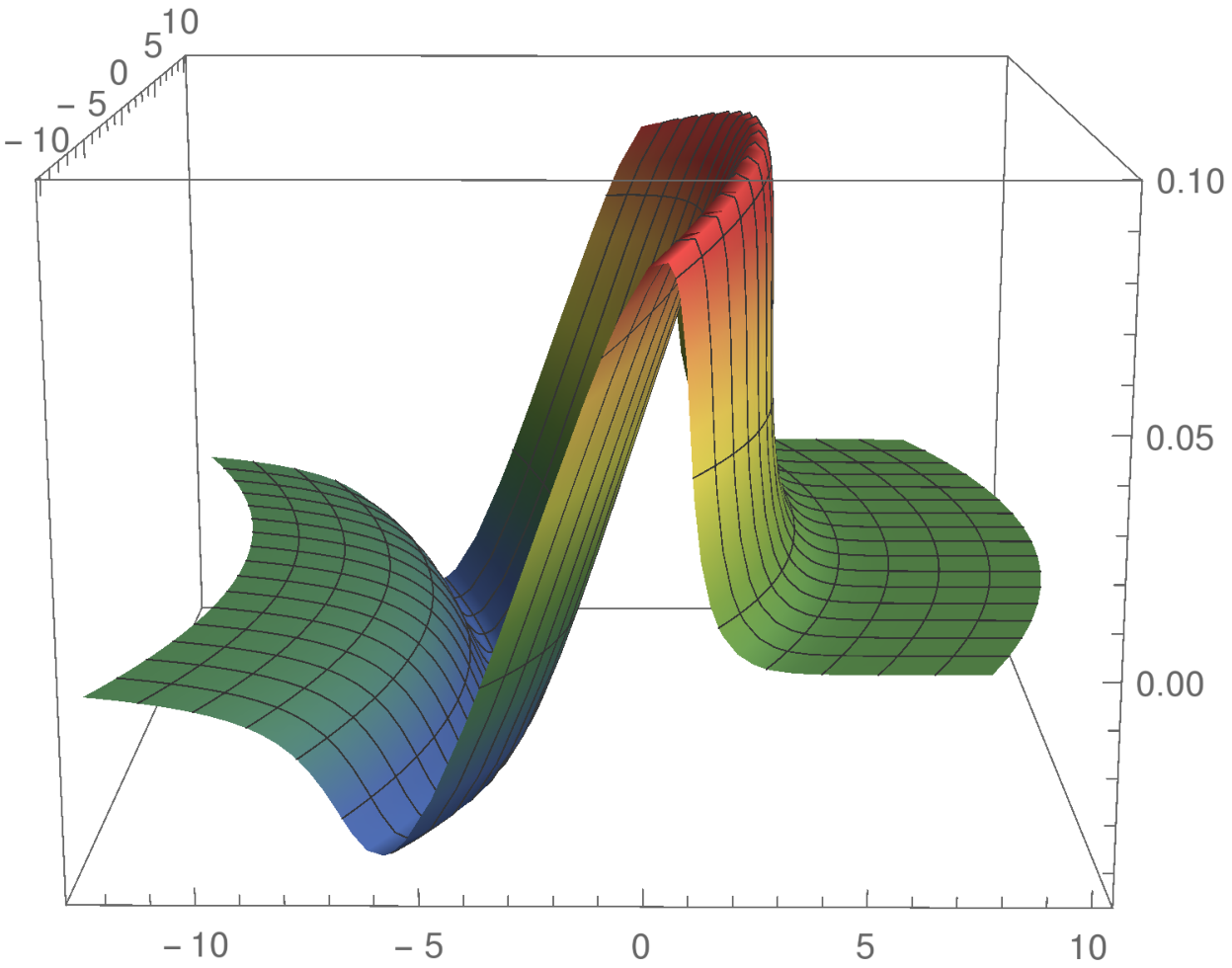}}
\end{center}

\noindent
Figure 4. Plotting of the analytic dKP solution $u$ in the physical variables $x,y,t$, for $G$ given by (\ref{G_gaussian_n=2}) (corresponding to the gaussian initial condition), $\eps=0.2$, at breaking.  
\vskip 10pt
For mdKP: 
\beq
\zeta_b=1.61,~\eta_b=0,~\xi_b=2.51,~\tau_b=14.31  
\eeq
and wave breaking takes place at $t_b=1.64\cdot  10^6$ in the point $(x_b,y_b)=(2.51, 0)$ on the parabola $x+1.52\cdot  10^{-7}y^2=2.51$, and at breaking the amplitude of the wave is $u=\eps {t_b}^{-1/2}G(\zeta_b,\eta_b)=1.95 \cdot  10^{-4}$ (see Figure 5). 
\vskip 10pt
\noindent
\begin{center}
\mbox{\epsfxsize=6cm \epsffile{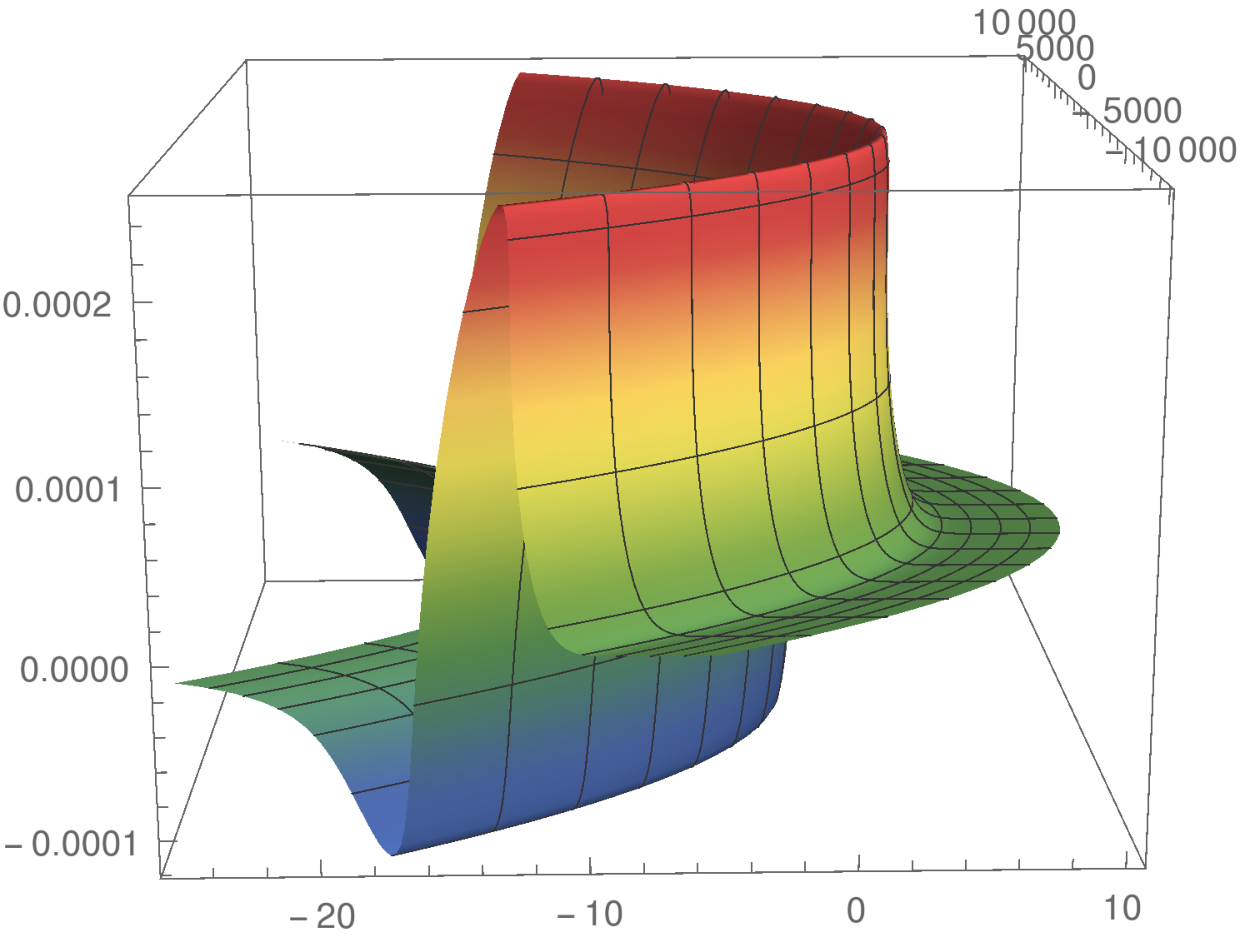}}
\end{center}

\noindent
Figure 5. Plotting of the analytic mdKP solution $u$ in the physical variables $x,y,t$, for $G$ given by (\ref{G_gaussian_n=2}) (corresponding to the gaussian initial condition), $\eps=0.2$, at breaking.
\vskip 5pt
Since, in the mdKP case, $t_b$ is very large and the wave amplitude at breaking is very small, we wonder if such a wave breaking could be actually detected numerically.

\vskip 5pt
Now we go into the analytic details. For the dKP case:  
\beq
\ba{l}
t'\equiv t-t_b\sim 2 t_b \tilde\tau, \ \ y'=y-y_b\simeq 2 t_b(\eta' +2\eta_b \tilde\tau +2\eta'\tilde\tau), \\ 
x' = x-x_b\sim \xi'-\eta_b y' -\frac{{y'}^2}{4 t_b}+\eta^2_b t' +\frac{\eta_b}{t_b}y' t',
\ea
\eeq 
and we choose $\tau'$ to be sufficiently small to have $|t' |,|y' | \ll 1$.

The dKP solution reads:
\beq
u(x,y,t)=\frac{\eps}{\sqrt{t}}G(\zeta_b+\zeta',\eta_b+\eta')=
\frac{\eps}{\sqrt{t}}\left(G+G_{\zeta}\zeta'+G_{\eta}\frac{y'}{2 t_b}+O\left(\frac{t'}{t_b}\right)\right),
\eeq
where $\zeta'$ is the solution of the cubic (\ref{cubic}), with:
\beq
\ba{l}
a=\frac{3F_{\zeta\zeta\eta}}{2 F_{\zeta\zeta\zeta}}\frac{y'}{t_b}, \ \ 
b=\frac{3}{F_{\zeta\zeta\zeta}}\left(F_{\zeta}\frac{t'}{t_b}+F_{\zeta\eta\eta}\frac{{y'}^2}{4 t^2_b} \right),\\
X=x'+\left(\eta_b -\frac{F_{\eta}}{2|F_{\zeta}|t_b}\right)y'-
\left[\eta^2_b+\left(\frac{F}{2}-\eta_b F_{\eta}\right)\frac{1}{|F_{\zeta}|t_b}\right]t'\\
+\left(1-\frac{F_{\eta\eta}}{2|F_{\zeta}|t_b}\right)\frac{{y'}^2}{4 t_b}+
\left[-\eta_b+\left(\frac{F_{\eta}}{2}+\eta_b F_{\eta\eta}\right)\frac{1}{2 |F_{\zeta}|t_b}\right]\frac{y' t'}{t_b}\\
-\frac{F_{\eta\eta\eta}}{48 |F_{\zeta}|}\frac{{y'}^3}{t^3_b}.
\ea
\eeq

The three-valued region (\ref{Delta=0}) after breaking becomes  
\beq
\label{Delta=0_dKP}
\ba{l}
\Big\{
x'+\left(\eta_b -\frac{F_{\eta}}{2|F_{\zeta}|t_b}\right)y'-
\left[\eta^2_b+\left(\frac{F}{2}-\eta_b F_{\eta}\right)\frac{1}{|F_{\zeta}|t_b}\right]t'\\
+\left(1-\frac{F_{\eta\eta}}{2|F_{\zeta}|t_b}\right)\frac{{y'}^2}{4 t_b}+
\left[-\eta_b+\left(\frac{F_{\eta}}{2|F_{\zeta}|t_b}+\frac{\eta_b F_{\eta\eta}}{|F_{\zeta}|}-
\frac{F_{\zeta\zeta\eta}}{2F_{\zeta\zeta\zeta} }\right)\frac{1}{2 t_b}\right]\frac{y' t'}{t_b}\\
+\frac{1}{6|F_{\zeta}|F^2_{\zeta\zeta\zeta}}\Big[-F_{\eta\eta\eta}F^2_{\zeta\zeta\zeta} + F_{\zeta\zeta\eta}(3F_{\zeta\zeta\zeta}F_{\zeta\eta\eta}-2F^2_{\zeta\zeta\eta})\Big]\frac{{y'}^3}{8 t^3_b}
\Big\}^2= \\
\frac{\alpha^3}{9\cdot 64 F^2_{\zeta} F^4_{\zeta\zeta\zeta} t^6_b} 
\left( \left(y'_{cusp}(t')\right)^2 - {y'}^2 \right)^3,
\ea
\eeq
where
\beq
y'_{cusp}(t')=2\sqrt{\frac{|F_{\zeta}|F_{\zeta\zeta\zeta}}{\alpha}}\sqrt{t_b t'} .
\eeq 
Equation (\ref{Delta=0_dKP}) describes a closed curve with two cusps in the $(x,y)$ - plane at the points
\beq\label{cusp dkp}
\vec x^{\pm}_{cusp}(t')=(x_b,y_b)+(x^{\pm '}_{cusp}(t'),\pm y'_{cusp}(t')) ,
\eeq
where
\beq\label{def_xicusps dkp}
\ba{l} 
x^{\pm '}_{cusp}(t')\sim 
\mp 2\eta_b \sqrt{\frac{|F_{\zeta}|F_{\zeta\zeta\zeta}}{\alpha}}\sqrt{t_b t'}
\pm F_{\eta}\sqrt{\frac{F_{\zeta\zeta\zeta}}{\alpha |F_{\zeta}|}}\sqrt{\frac{t'}{t_b}}\\
+\left(\eta^2_b -\frac{|F_{\zeta}|F_{\zeta\zeta\zeta}}{\alpha }\right)t'+
\left(\frac{F}{|F_{\zeta}|}+\frac{F_{\eta\eta}F_{\zeta\zeta\zeta}}{\alpha}\right)\frac{t'}{2 t_b}
\pm 2\eta_b \sqrt{\frac{|F_{\zeta}|F_{\zeta\zeta\zeta}}{\alpha}}\frac{{t'}^{\frac{3}{2}}}{\sqrt{t_b}}
+O({\tilde\tau}^{3/2})
\ea
\eeq
At last, the analogue of (\ref{longitudilal width}) reads
\beq\label{longitudilal width dkp}
\Delta x'\le \frac{2}{3}\sqrt{\frac{|F_{\zeta}|}{F_{\zeta\zeta\zeta}}}\left(\frac{t'}{t_b} \right)^{\frac{3}{2}};
\eeq
it follows that the longitudinal width $\Delta_{long}$ and the transversal width $\Delta_{transv}$ of the multivalued region are respectively 
\beq
\ba{l}
\Delta_{long}=O\left(\left(\frac{t'}{t_b} \right)^{\frac{3}{2}}\right), \ \ 
\Delta_{transv}=O\left(\left(t_b t'\right)^{\frac{1}{2}}\right), \\ 
\Rightarrow \ \ \frac{\Delta_{long}}{\Delta_{transv}}=O\left(\frac{t'}{t^2_b} \right).
\ea
\eeq

For the mdKP case:
\beq
\ba{l}
t'=t-t_b\sim t_b\tau', \ \ y'=y-y_b\sim 2t_b (\eta'+\eta_b\frac{t'}{t_b}+\frac{\eta' t'}{t_b}), \\
x'=x-x_b\sim \xi' -\eta_b y' +\eta^2_b t'-\frac{{y'}^2}{4 t_b}+\eta_b\frac{y' t'}{t_b},
\ea
\eeq
and we choose $\tau'$ sufficiently small to have $|t'|,~|y'|\ll 1$.

The mdKP solution reads:
\beq
u(x,y,t)=\frac{\eps}{\sqrt{t}}G(\zeta_b+\zeta',\eta_b+\eta')=
\frac{\eps}{\sqrt{t}}\left(G+G_{\zeta}\zeta'+G_{\eta}\frac{y'}{2 t_b}+O\left(\frac{t'}{t_b\log t_b}\right)\right),
\eeq
where $\zeta'$ is the solution of the cubic (\ref{cubic}), with:
\beq
\ba{l}
a=\frac{3F_{\zeta\zeta\eta}}{2 F_{\zeta\zeta\zeta}}\frac{y'}{t_b}, \ \ 
b=\frac{3}{F_{\zeta\zeta\zeta}}\left(2F_{\zeta}\frac{t'}{t_b\log t_b}+F_{\zeta\eta\eta}\frac{{y'}^2}{4 t^2_b} \right),\\
X=x'+\left(\eta_b -\frac{F_{\eta}}{2|F_{\zeta}|t_b}\right)y'-
\left[\eta^2_b+\left(\frac{F}{\log t_b}-\eta_b F_{\eta}\right)\frac{1}{|F_{\zeta}|t_b}\right]t'\\
+\left(1-\frac{F_{\eta\eta}}{2|F_{\zeta}|t_b}\right)\frac{{y'}^2}{4 t_b}+
\left[-\eta_b+\left(\frac{F_{\eta}}{|F_{\zeta}|}(1-\frac{1}{\log t_b})+\frac{\eta_b F_{\eta\eta}}{|F_{\zeta}|}\right)\frac{1}{2 t_b}\right]\frac{y' t'}{t_b}\\
-\frac{F_{\eta\eta\eta}}{48 |F_{\zeta}|}\frac{{y'}^3}{t^3_b}.
\ea
\eeq 

The three-valued region (\ref{Delta=0}) after breaking becomes  
\beq
\label{Delta=0_mdKP}
\ba{l}
\Big\{
x'+\left(\eta_b -\frac{F_{\eta}}{2|F_{\zeta}|t_b}\right)y'-
\left[\eta^2_b+\left(\frac{F}{\log t_b}-\eta_b F_{\eta}\right)\frac{1}{|F_{\zeta}|t_b}\right]t'\\
+\left(1-\frac{F_{\eta\eta}}{2|F_{\zeta}|t_b}\right)\frac{{y'}^2}{4 t_b}+
\left[-\eta_b+\left(\frac{F_{\eta}}{|F_{\zeta}|}(1-\frac{1}{\log t_b})+\frac{\eta_b F_{\eta\eta}}{|F_{\zeta}|}-
\frac{F_{\zeta\zeta\eta}}{F_{\zeta\zeta\zeta}\log t_b }\right)\frac{1}{2 t_b}\right]\frac{y' t'}{t_b}\\
+\frac{1}{6|F_{\zeta}|F^2_{\zeta\zeta\zeta}}\Big[-F_{\eta\eta\eta}F^2_{\zeta\zeta\zeta} + F_{\zeta\zeta\eta}(3F_{\zeta\zeta\zeta}F_{\zeta\eta\eta}-2F^2_{\zeta\zeta\eta})\Big]\frac{{y'}^3}{8 t^3_b}
\Big\}^2= \\
\frac{\alpha^3}{9\cdot 64 F^2_{\zeta} F^4_{\zeta\zeta\zeta} t^6_b} 
\left( \left(y'_{cusp}(t')\right)^2 - {y'}^2 \right)^3,
\ea
\eeq
where
\beq
y'_{cusp}(t')=2\sqrt{\frac{2|F_{\zeta}|F_{\zeta\zeta\zeta}}{\alpha}}\sqrt{\frac{t_b t'}{\log t_b}} .
\eeq 
Equation (\ref{Delta=0_mdKP}) describes a closed curve with two cusps in the $(x,y)$ - plane at the points
\beq\label{cusp mdkp}
\vec x^{\pm}_{cusp}(t')=(x_b,y_b)+(x^{\pm '}_{cusp}(t'),\pm y'_{cusp}(t')) ,
\eeq
where
\beq\label{def_xicusps dkp}
\ba{l} 
x^{\pm '}_{cusp}(t')\sim 
\mp 2\eta_b \sqrt{\frac{|F_{\zeta}|F_{\zeta\zeta\zeta}}{\alpha}}\sqrt{\frac{t_b t'}{\log t_b}}
\pm F_{\eta}\sqrt{\frac{F_{\zeta\zeta\zeta}}{\alpha |F_{\zeta}|}}\sqrt{\frac{t'}{t_b\log t_b}}\\
+\left(\eta^2_b -\frac{|F_{\zeta}|F_{\zeta\zeta\zeta}}{\alpha \log t_b}\right)t'+
\left(\frac{F}{|F_{\zeta}|}+\frac{F_{\eta\eta}F_{\zeta\zeta\zeta}}{\alpha}\right)\frac{t'}{t_b\log t_b}
\pm 2\eta_b \sqrt{\frac{|F_{\zeta}|F_{\zeta\zeta\zeta}}{\alpha}}\frac{{t'}^{\frac{3}{2}}}{\sqrt{t_b\log t_b}}\\
+O({\tilde\tau}^{3/2})
\ea
\eeq
At last, the analogue of (\ref{longitudilal width}) reads
\beq\label{longitudilal width mdkp}
\Delta x'\le \frac{4\sqrt{2}}{3}\sqrt{\frac{|F_{\zeta}|}{F_{\zeta\zeta\zeta}}}\left(\frac{t'}{t_b\log t_b} \right)^{\frac{3}{2}};
\eeq
it follows that the longitudinal width $\Delta_{long}$ and the transversal width $\Delta_{transv}$ of the multivalued region are respectively 
\beq
\ba{l}
\Delta_{long}=O\left(\left(\frac{t'}{t_b\log t_b} \right)^{\frac{3}{2}}\right), \ \ 
\Delta_{transv}=O\left(\left(\frac{t_b t'}{\log t_b}\right)^{\frac{1}{2}}\right), \\ 
\Rightarrow \ \ \frac{\Delta_{long}}{\Delta_{transv}}=O\left(\frac{t'}{t^2_b\log t_b} \right).
\ea
\eeq


\end{document}